\documentclass{rsauthor}	

\usepackage{lscape}
\usepackage{afterpage}

\usepackage[sectionbib]{natbib}

\bibpunct[, ]{(}{)}{;}{a}{}{,}

\let\mycite\citep
\let\myciteasnoun\citet

\renewcommand\ge\geqslant
\renewcommand\geq\geqslant
\renewcommand\le\leqslant
\renewcommand\leq\leqslant
\renewcommand\propto\varpropto

\newcommand{\defeq}{:=}
\newcommand{\eqdef}{\mathrel{=:}}

\theoremstyle{definition}
\newtheorem*{example*}{Example}
\newtheorem{example}{Example}[section]

\jname{Proc.\ R. Soc.\ A}

\begin{document}

\title[Integration of three-dimensional Lotka--Volterra systems]{The integration of three-dimensional\\ Lotka--Volterra systems}
\author[R. S. Maier]{\sc By Robert S. Maier${}^*$\footnote{${}^*$rsm@math.arizona.edu}}
\address{Departments of Mathematics and Physics, University of Arizona,\\ Tucson, AZ 85721, USA}
\abstract{The general solutions of many three-dimensional Lotka--Volterra
  systems, previously known to be at~least partially integrable, are
  constructed with the aid of special functions.  Examples include certain
  ABC and May--Leonard systems.  The special functions used are incomplete
  beta and elliptic functions.  In some cases the solution is parametric,
  with the independent and dependent variables expressed as functions of a
  `new time' variable.  This auxiliary variable satisfies a nonlinear
  third-order differential equation of a generalized Schwarzian type, and
  results of Carton-LeBrun on such equations are exploited.  Several
  difficult Lotka--Volterra systems are successfully integrated in~terms of
  Painlev\'e transcendents.  An appendix on incomplete beta functions is
  included.}

\makeatletter
\markboth{\@shortauthor\hfil}{\hfil\@shorttitle}
\makeatother

\maketitle

\section{Introduction}
\label{sec:1}
\subsection{Background}
\label{subsec:1a}

Models of Lotka--Volterra type occur frequently in the physical and
engineering sciences, as well as the biological.  In any such model there
is a $d$\nobreakdash-dimensional state vector $x=(x_1,\dots,x_d)$, a
function of time~$\tau$, which satisfies an autonomous system of ordinary
differential equations (ODE's) of the form
\begin{equation}
\label{eq:LV}
  \dot x_i = x_i\Biggl( a_{i0} + \sum_{j=1}^d a_{ij}x_j \Biggr),\qquad
  i=1,\dots,d.
\end{equation}
The overdot signifies ${\rm d}/{\rm d}\tau$.  By letting $x_0\equiv1$ and
$a_{0i}\equiv0$ one can optionally rewrite this as $\dot { x}_i =
x_i\sum_{j=0}^d a_{ij} x_j$, $i=0,\dots,d$.  There is a paucity
of closed-form solutions: when $d\ge3$ and even when $d=2$, the
system~(\ref{eq:LV}) is usually integrated numerically rather than
symbolically.

In most applications $x_i\ge0$ for all~$i$, though it may be useful to
allow the~$x_i$ and even~$\tau$ to be complex.  Such models were first
introduced in population ecology \mycite{Kot2001,May2001}, where $x$~lies
in the non-negative orthant, and the terminology reflects this: $a_{i0}$~is
the growth rate of the $i$'th `species,' and depending on the signs of the
elements of the interaction matrix ${\sf A}=(a_{ij})_{i,j=1}^d$, one speaks
of predation, competition or mutualism.  If ${a_{i0}>0}$ and ${a_{ii}<0}$,
in the absence of the other $d-\nobreak 1$ species the $i$'th species will
grow logistically, but the behaviour of $x=x(\tau)$ in the interior of the
orthant may be much more complicated.

Modelling by small-$d$ Lotka--Volterra systems has occurred many times in
physics.  In laser physics this ranges from the initial treatment of
multimode coupling \mycite{Lamb64} to Raman amplification in an optical
fibre \mycite{Castella2004}, where $\tau$~is the distance along the fibre.
Other physical applications include the modelling of the (integrable)
interaction of Langmuir waves in plasmas \mycite{Qin2011} and more chaotic
phenomena: the Kueppers--Lortz instability in a rotating fluid
\mycite{Busse83,SanMiguel2000}, for which $d=3$, and the Boltzmann dynamics
of $d$~rarefied, spatially homogeneous gases in a background medium
\mycite{Jenks69,Lupini88}.  Several integrable Lotka--Volterra systems with
fairly small~$d$, of mathematical interest, have been obtained as spatial
discretizations of the Korteweg--\allowbreak{de}~Vries equation, or as
generalizations of same \mycite{Bogoyavlensky88,Bogoyavlenskij2008}.

In population biology, small\nobreakdash-$d$ Lotka--Volterra systems
include the classical $d=2$ predator--prey model of Volterra and Lotka,
well-known though structurally unstable; the famous May--Leonard model of
$d=3$ cyclically competing species with equal growth rates, which has a
stable limit cycle \mycite{May75}; and generalizations that model the
asymmetric competition of $d$~species in a chemostat
\mycite{Wolkowicz2006,Ajbar2012}.  There is also a literature on
evolutionarily stable strategies in game dynamics, modelled by differential
systems equivalent to Lotka--Volterra ones
\mycite{Hofbauer88,Hofbauer2003}.  Lotka--Volterra systems with $d\ge3$ may
display chaotic behaviour, a fact first noted in ecological modelling
\mycite{Klebanoff94,Vano2006}.

Many nonquadratic nonlinear differential systems can be transformed
to~(\ref{eq:LV}) by changes of variable
\mycite{Peschel86,Brenig89,Gouze92,Hernandez97}.  Complex chemical
reactions with mass-action kinetics provide examples.  For special
stoichiometries, spatially homogeneous species concentrations evolve in a
Lotka--Volterra way \mycite{Erdi89a,Murza2010}.  Furthermore, mass-action
systems with nonquadratic polynomial nonlinearities, which can be obtained
from (formal) chemical reactions in more than one way, can often be
transformed to the Lotka--Volterra form.

Several elementary changes of variable deserve mention, for any
$d$-dimensional system of the form~(\ref{eq:LV}).  First, by scaling~$\tau$
one can scale the column vector of growth rates and the interaction
matrix~$\mathsf{A}$.  Additionally, by scaling $x_1,\dots,x_d$
independently one can scale independently the $d$~columns of~$\mathsf{A}$.
Thus the parameter space has dimensionality not $d^2+\nobreak d$ but
$d^2\nobreak-1$.

If the growth rates $a_{10},\dots,a_{d0}$ are \emph{equal} (to~$a_{*0}$,
say), then in~terms of a transformed time $\tau'\defeq {\rm e}^{a_{*0}\tau}$ the
transformed variables $x_i'\defeq {\rm e}^{-a_{*0}\tau}x_i$ will satisfy a system
of the form~(\ref{eq:LV}) but with zero growth rates.  Furthermore
\mycite{Plank99}, if $x_1,\dots,x_d$ satisfy (\ref{eq:LV}) with zero growth
rates then the ratios $(\tilde x_1,\dots,\tilde x_{d-1})\defeq\allowbreak
(x_1/x_d,\dots,x_{d-1}/x_d)$ will satisfy a $(d-\nobreak1)$-dimensional
system resembling~(\ref{eq:LV}), but with growth rates $\tilde a_{i0}\defeq
a_{id}-\nobreak a_{dd}$ and interactions $\tilde a_{ij}\defeq \allowbreak
a_{ij}-\nobreak a_{dj}$.  Thus Lotka--Volterra systems with equal growth
rates, for which the parameter space dimensionality is effectively
$d^2-\nobreak d$, should be peculiarly amenable to analysis; though even
they have usually been studied not symbolically but numerically.

\smallskip
The object of this paper is the construction of general solutions
$x=x(\tau)$ of systems of the form~(\ref{eq:LV}), with the aid of special
functions.  The focus is on $d=3$ systems, though ancillary results on
integrable $d=2$ systems are obtained first.  The special functions
employed are the incomplete beta function and its inverse, which sometimes
reduces to an elementary or elliptic function.  For many $d=3$ systems
known to be at~least partially integrable, general solutions are
constructed for the first time, with unexpected ease.  By default, each of
$x,\tau$ is expressed as a function of an auxiliary variable~$t$, a~`new
time.'  A minor but interesting example is the fully (rather than merely
cyclically) symmetric May--Leonard model.

Most of the integrated systems have equal growth rates, but not all.
Several systems with unequal ones which have the Painlev\'e property, that
each solution $x=x(\tau)$ can be extended analytically to a one-valued
function on the complex $\tau$\nobreakdash-plane, are integrated with the
aid of Painlev\'e transcendents.

Many of the interesting examples of general solutions $x=x(\tau)$ that are
presented are made possible by results of \myciteasnoun{CartonLeBrun69c}.
She classified all nonlinear third-order ODE's of a certain type that have
the Painlev\'e property; and the ODE satisfied by the new time $t=t(\tau)$
used here was fortuitously included.  She also integrated many such ODE's
with the aid of elliptic functions.  The examples in \S\S\,\ref{sec:3}
and~\ref{sec:4} may serve to reawaken interest in her work.

This paper includes an appendix on closed-form expressions for inverse
incomplete beta functions, which should be of independent interest.

\subsection{Previous results, reformulated}
\label{subsec:1b}

The following is a summary of symbolic (non-numerical) results on
small\nobreakdash-$d$ Lotka--Volterra systems, which provides the context
for the new results.

Lotka--Volterra solutions cannot always be expressed in closed form,
i.e.\ in~terms of elementary or `known' functions, even if the behaviour of
the solutions is nonchaotic; or so it is believed.  An example is the
original system of Volterra and Lotka, with $d=2$ and no self-interactions
($a_{11}=a_{22}=0$).  If the growth rates are equal it has a general
solution in~terms of elementary functions \mycite{Varma77}.  If the rates
are unequal it is nonetheless integrable (see below) and hence nonchaotic,
but this does not imply the existence of a useful closed form for the
general solution.  The case of unequal growth rates when $d=2$ has in~fact
been used as a test-bed for geometric numerical integration schemes
\mycite{Hairer2006}.

Owing to the difficulty of constructing explicit solutions of systems of
the form~(\ref{eq:LV}), there has been much work on the simpler problem of
finding constants of the motion, i.e., first integrals.  Any such is a
function $I=I(x_1,\dots,x_d)$, preferably one-valued and well-behaved,
obeying ${\dot I=0}$ on any system trajectory.  (By the overdot the total
time derivative is meant.)  If such an~$I$ exists, any trajectory is
confined to a $(d-\nobreak1)$-dimensional surface
$I\equiv{\textrm{const}}$; which, e.g., facilitates the study of long-time
behaviour.  If there are $d-\nobreak m$ functionally independent first
integrals then motion is confined to a $m$\nobreakdash-dimensional surface,
obtained as an intersection.  When $m=1$ the system is completely
integrable (its general solution can be `reduced to quadratures,' though
not necessarily usefully).  When $m=2$ it is partially integrable and is
at~least nonchaotic.

For $d=2,3$, restrictions on the Lotka--Volterra parameters
$\{a_{ij}\}_{i=1,j=0}^d$ that imply the existence of a single closed-form
first integral have been investigated by many authors.  In most cases the
resulting first integral is of the form
\begin{equation}
\label{eq:IDP}
  I=\prod\nolimits_{k=1}^r \left|f_k\right|^{l_k},
\end{equation}
with each $f_k=f_k(x_1,\dots,x_d)$ a polynomial (usually though not always
of degree~1 with constant term allowed).  Results along this line include
those of Cair\'o and collaborators
\mycite{Cairo92,Cairo1999,Cairo2000a,Cairo2000b}, who in many cases used
the classical technique of Darboux polynomials \mycite{Goriely2001}; and
results based on an integrating factor technique \mycite{Saputra2010}.
When $d=3$ and a single first integral of a form similar to~(\ref{eq:IDP})
exists, there has been some work on the construction of a second,
functionally independent first integral of a more complicated and less
algebraic form \mycite{Grammaticos90,Goriely92,Gao2000,Bustamante2003}.

\afterpage{\clearpage\rm
\begin{landscape}
\vfill
\begin{table}
  \caption{For $d=2$, cases when $\le2$~restrictions yield a Darboux
    polynomial~$f$ and a first integral: $\left|x_1\right|^{l_1}\left|x_2\right|^{l_2}\left|f\right|^{l_3}$.}
  \begin{center}
  \begin{tabular}{llll}
    \hline
    case & restrictions & DP $f$ & cofactor of~$f$ \\
    \hline
    \hline
    $\textrm{I}$ & $\det(a_0,a_2) = \det(a_1,a_0) = 0$ & none needed & --- \\
    $\textrm{I}_i'$ & $\det(a_i,a_j) = \det(a_0,a_i) = 0$ & none needed & --- \\
    \hline
    $\textrm{II}$ & $a_{10}=a_{20}=0 $ & $(a_{11}-a_{21})x_1 - (a_{22}-a_{12})x_2$ & $a_{11}\,x_1 + a_{22}\,x_2$ \\
    $\textrm{II}_i'$ & $a_{ij}=a_{jj}=0 $ & $a_{i0}+a_{ii}\,x_i$ & $a_{ii}\,x_i$ \\
    \hline
    $\textrm{III}$ & $r_{012}=0$ & $a_{10}a_{20} + a_{20}a_{11}\,x_1 +  a_{10} a_{22} \,x_2 $  & $a_{11}\,x_1+a_{22}\,x_2$ \\
    \hline
  \end{tabular}
  \end{center}
  \label{tab:2tab}
\end{table}
\vfill
\begin{table}
  \caption{For $d=3$, cases when $\le3$~restrictions yield a Darboux
    polynomial~$f$ and a first integral:
    $\left|x_1\right|^{l_1}\left|x_2\right|^{l_2}\left|x_3\right|^{l_3}\left|f\right|^{l_4}$.}
  \begin{center}
  \begin{tabular}{llll}
    \hline
    case & restrictions & DP $f$ & cofactor of~$f$ \\
    \hline
    \hline
    $\textrm{I}$ & $\det(a_0,a_2,a_3)=\det(a_1,a_0,a_3)$ & none needed & --- \\
    & $\qquad\qquad=\det(a_1,a_2,a_0)=0$ && \\
    $\textrm{I}_i'$ & $\det(a_i,a_j,a_k)=\det(a_0,a_i,a_k)$  & none needed & --- \\
    & $\qquad\qquad=\det(a_0,a_j,a_i)=0$  && \\
    \hline
    $\textrm{II}_i$ & $a_{j0}=a_{k0}\eqdef a_{*0}$, & $(a_{jj}-a_{kj})x_j - (a_{kk}-a_{jk})x_k$ & $a_{*0} + a_{*i}\,x_i + a_{jj}\,x_j + a_{kk}\,x_k $  \\
    & $a_{ji}=a_{ki}\eqdef a_{*i}$, &  \\
    & $a_{i0}a_{*i}=a_{*0}a_{ii}$ && \\
    $\textrm{II}_i'$ & $a_{ij}=a_{ik}=0$, & $a_{i0} + a_{ii}\,x_i $ & $a_{ii}\,x_i$\\
    & $a_{jj}a_{kk}=a_{jk}a_{kj}$ && \\
    \hline
    $\textrm{III}_i$ & $r_{0jk}=0$, $a_{ji}=a_{ki}=0$ & $a_{j0}a_{k0} + a_{k0}a_{jj}\,x_j + a_{j0}a_{kk}\,x_k $ & $a_{jj}\,x_j +    a_{kk}\,x_k$ \\
    $\textrm{III}'$ & $r_{123} = 0$, $a_{10}=a_{20}=a_{30}\eqdef a_{*0}$ & $(a_{ji}-a_{ii})(a_{ki}-a_{ii})x_i $  & $a_{*0} + a_{11}\,x_1 + a_{22}\,x_2 + a_{33}\,x_3$ \\
    & & $\quad {}- (a_{ij}-a_{jj})(a_{ki}-a_{ii})x_j $  &  \\
    & & $\quad {}- (a_{ik}-a_{kk})(a_{ji}-a_{ii})x_k $  &  \\
    \hline
  \end{tabular}
  \end{center}
  \label{tab:3tab}
\end{table}
\end{landscape}
}                               

The Darboux polynomial (DP) concept is fundamental.  A~DP
$f=f(x_1,\dots,x_d)$ is a polynomial for which $\dot f=Kf$ along any system
trajectory, where the `cofactor'~$K=K(x_1,\dots,x_d)$ is necessarily also a
polynomial, of degree~1 with constant term allowed.  The algebraic surface
$f=0$ is thus invariant under the Lotka--Volterra flow.  If the cofactors
of DP's $f_1,\dots f_r$ are linearly dependent, some product of the
form~(\ref{eq:IDP}) will be a first integral: generically, a non-trivial
one.  Finding a first integral, including the needed parametric
restrictions, is facilitated by the coordinate functions $x_1,\dots,x_d$
being DP's, associated to the invariant hyperplanes $x_i=0$, $i=1,\dots,d$,
and having respective cofactors $K_i=\sum_{j=0}^d a_{ij}x_j$.  To construct
a first integral a single additional DP~$f$ will suffice, if one
sufficiently restricts parameters to ensure dependence of the cofactors of
$x_1,\dots,x_d,f$.

Tables \ref{tab:2tab} and~\ref{tab:3tab} are a distillation of the results
of the preceding authors on degree\nobreakdash-1 DP's~$f$ that yield first
integrals of the form
$\left|x_1\right|^{l_1}\!\!\dots\left|x_d\right|^{l_d}\left|f\right|^{l_{d+1}}$.
In the two tables, $i,j$ and $i,j,k$ are arbitrary permutations of $1,2$
and~$1,2,3$, respectively; and $f=0$ is an invariant line, resp.\ plane.
For both $d=2$ and~$3$, the cases when $\le d$ restrictions on the
parameters yield a first integral of this form split into three types:
I,~II and~III\null.  (The numbering here differs from that of
\myciteasnoun{Cairo92} but agrees with that of \myciteasnoun{Hua96}.)  The
primed cases can be viewed as projectively transformed versions of the
unprimed ones.  For instance, the $d=2$ case~$\textrm{II}_i'$ is related
to~II by $(\tilde x_0,\tilde x_i, \tilde x_j)=(x_j,x_i,x_0)/x_j$.

The type-I cases are degenerate: the first integral~$I$ does not require
any additional DP~$f$ and is merely of the form $\prod_{i=1}^d
\left|x_i\right|^{l_i}$.  In fact, in each type\nobreakdash-I case the
restrictions on the parameters, expressed in~terms of the $d+\nobreak1$
column vectors $a_j\defeq(a_{ij})_{i=1}^d$, $j=0,\dots,d$, imply that the
matrix $\mathsf{A}=(a_{ij})_{i=1,j=1}^d$ must be singular, i.e.,
$\det\mathsf{A}=0$.  Lotka--Volterra models with singular~$\mathsf A$
(e.g., ones in which $\mathsf{A}$~is antisymmetric and $d$~is odd) may have
physical applications, though their relevance to population ecology has
been strongly questioned \mycite{MaynardSmith74,May2001}.  But they will
not be considered further here.

For $d=2$, in case~II (which is defined by $a_{10}=a_{20}=0$, i.e., by the
growth rates of the two species being zero), the resulting first integral
is easily seen to be
\begin{equation}
  \label{eq:1stintegral2}
  I = \left|x_1\right|^{a_{22}(a_{21}-a_{11})}\, \left|x_2\right|^{a_{11}(a_{12}-a_{22})}
  \,\left|(a_{11}-a_{21})x_1 - (a_{22}-a_{12})x_2\right|^\Delta,
\end{equation}
where $\Delta\defeq\det\mathsf{A}$.  For $d=3$, the similar
case~$\textrm{II}_i$ is defined by $a_{j0}=a_{k0}\eqdef a_{*0}$, i.e., by
the two species other than species~$i$ having equal growth rates, and by
$a_{ji}=a_{ki}\eqdef a_{*i}$, i.e., by their having equal effects on
species~$i$.  A~third condition $a_{i0}a_{*i}=a_{*0}a_{ii}$ must also be
satisfied.  (Note that it will be satisfied if all growth rates are zero.)
The resulting first integral is
\begin{multline}
  \label{eq:1stintegral3}
    I = \left|x_i\right|^{a_{*i}(a_{kj}-a_{jj})(a_{jk}-a_{kk})}\,
    \left|x_j\right|^{(a_{ii}a_{kk}-a_{ik}a_{ki})(a_{kj}-a_{jj})}\,
    \left|x_k\right|^{(a_{ii}a_{jj}-a_{ij}a_{ji})(a_{jk}-a_{kk})}\\[-2.5pt]
    {}\times \left|(a_{jj}-a_{kj})x_j - (a_{kk}-a_{jk})x_k\right|^\Delta.
\end{multline}
The $d=3$ expression (\ref{eq:1stintegral3}) reduces to a power of the
$d=2$ first integral~(\ref{eq:1stintegral2}) when $(i,j,k)=(3,1,2)$ and the
third species decouples: $a_{*i}=0$.

The tables also list restrictions of type~III, which are expressed in~terms
of the quantities
\begin{equation}
  r_{ijk}\defeq
  (a_{ji}-a_{ii})(a_{kj}-a_{jj})(a_{ik}-a_{kk})
  + (a_{ki}-a_{ii})(a_{ij}-a_{jj})(a_{jk}-a_{kk}).
\end{equation}
But the present paper will focus entirely on $d=2$ and especially $d=3$
systems that satisfy the less complicated restrictions of type~II,
resp.\ type~$\textrm{II}_i$.

It must be stressed that Lotka--Volterra systems with $d=2,3$ and no
self-interactions ($a_{ii}=0$ for $i=1,\dots,d$) do not fit well into the
DP framework of tables \ref{tab:2tab} and~\ref{tab:3tab}.  One example is
the classical $d=2$ model, which has the two restrictions $a_{11}=a_{22}=0$
and is quite anomalous.  It is integrable with first integral
\begin{equation}
\label{eq:origI}
  I = \left|x_1\right|^{a_{20}}\left|x_2\right|^{-a_{10}} [\exp  (a_{21}x_1-a_{12}x_2)] \eqdef
  {\prod\nolimits}_{i=1}^3\left|f_i\right|^{l_i}.
\end{equation}
A~framework into which this model fits is only now being developed
\mycite{Llibre2009a}.  Each factor $f_i$ in~(\ref{eq:origI}) satisfies
$\dot f_i=K_if_i$ for some degree\nobreakdash-1 polynomial cofactor~$K_i$
and is therefore a DP or generalized~DP, sometimes called a `second
integral' \mycite{Goriely2001}.  To understand~(\ref{eq:origI}) one needs
the concept of the \emph{multiplicity} of an invariant line (or algebraic
curve) of a polynomial vector field \mycite{Christopher2007}.  The DP's
$f_1=x_1$, $f_2=x_2$ are associated to the invariant lines $x_1=0$,
$x_2=0$, which by default are not multiple.  The exponential factor~$f_3$
in~(\ref{eq:origI}) comes from the line `at~infinity,' which like $x_1=0$
and $x_2=0$ is invariant in any Lotka--Volterra model but which when
$a_{11}=a_{22}=0$ has a nontrivial (double) multiplicity.  It therefore
gives rise to an extra (generalized)~DP, namely~$f_3$.

The extent to which imposing parametric restrictions on a $d=2$
Lotka--Volterra system can produce single or multiple invariant lines,
thereby engendering DP's or generalized DP's, is now fully understood
\mycite{Schlomiuk2010}.  The number of invariant lines, counted with
multiplicity, can if finite be as large as six (including the one at
infinity).  The line configuration 4.18 of Schlomiuk \& Vulpe corresponds
to the classical model ($a_{11}=a_{22}=0$), and their 4.5 and~4.1 to the
types II and~III of table~\ref{tab:2tab}.  But an integrability study of
their many other configurations, most defined by severe parametric
restrictions and many including multiple invariant lines, remains to be
carried~out.  A $d=3$ counterpart to their ${d=2}$ analysis, which is
lengthy, is not yet available.  The many possible configurations of
invariant planes have not been fully classified, though partial results
have been obtained \mycite{Cairo2000a,Saputra2010}.

There is already a large literature on the integrability properties of
$d=3$ Lotka--Volterra systems without self-interactions
\mycite{Grammaticos90,Cairo2000b,Moulin2001,Moulin2004}.  They are called
ABC or $A_1A_2A_3$ systems, since if $a_{11}=a_{22}=a_{33}=0$ and
$a_{ij}a_{jk}a_{ki}\neq0$ for some permutation $ijk$ of~$123$, permuting
species and scaling columns will yield an interaction matrix
\begin{equation}
\label{eq:AABC}
  \mathsf{A}=(a_{ij})_{i,j=1}^3=\left(
  \begin{array}{ccc}
    0  &  A_2  &  1 \\
    1  &  0    & A_3 \\
    A_1 &  1   & 0 
  \end{array}
\right).
\end{equation}
If an ABC system satisfies $A_1A_2A_3=-1$ or $A_i=1$ for some~$i$, and
growth rates are suitably constrained, a first integral for~it can be
constructed from degree\nobreakdash-1 DP's.  This is because such a system
is of type~I or type~II, as defined in table~\ref{tab:3tab}.  But there are
exotic ABC systems with first integrals based on DP's of degree~2 or
greater, or on generalized DP's.  Being tightly restricted parametrically,
they are not covered by table~\ref{tab:3tab}.  Some of these systems were
first obtained by a Painlev\'e analysis \mycite{Bountis84}.  To some of
these exotic ABC systems exotic $d=2$ systems are associated, as any $d=3$
system with equal growth rates can be reduced to a $d=2$ system, typically
with unequal growth rates \mycite{Cairo2003}.

A $d=3$ system with even tighter parametric restrictions is the
May--Leonard model of cyclic competition \mycite{May75}.  This is a system
with equal growth rates (reducible to zero by a change of variables), and
with
\begin{equation}
\label{eq:MLmatrix}
  \mathsf{A}=(a_{ij})_{i,j=1}^3=\left(
  \begin{array}{ccc}
    -1  &  -\alpha  &  -\beta \\
    -\beta  &  -1    & -\alpha \\
    -\alpha &  -\beta   & -1 
  \end{array}
\right).
\end{equation}
If, e.g., $\alpha+\beta=-1$ or $\alpha=\beta=1$, then $\det\mathsf{A}=0$
and this system is of type~I in the sense of table~\ref{tab:3tab}.  If
$\alpha=\beta$ (so that it is fully rather than cyclically symmetric), then
it is of type~II, and specifically it is case~${\textrm{II}_i}$ for
each~$i$.  Hence it is completely integrable: for each permutation $ijk$
of~$123$, the expression~(\ref{eq:1stintegral3}) will be a first integral
(cf.~\myciteasnoun{Strelcyn88}).  Normalized, this is
\begin{equation}
\label{eq:strelcyn}
  I_i = \left|x_i\right|^{-\alpha} \left|x_j\right|^{\alpha+1} \left|x_k\right|^{\alpha+1} \left|x_j-x_k\right|^{-1-2\alpha}.
\end{equation}
Other first integrals are known \mycite{Llibre2011,Tudoran2012}.



The relation between the integrability (partial or complete) of a
differential system and its having the Painlev\'e property~(PP), i.e., the
property that each of its solutions $x=x(\tau)$ has a one-valued
continuation to the complex $\tau$\nobreakdash-plane (without branch points
of any order), is a bit murky.  It is widely believed that the integration
of any system with the~PP should reduce to quadratures; or if not, that the
solutions of the system should be expressible in~terms of `known'
functions, such as the Painlev\'e transcendents that arise as solutions of
second-order ODE's with the~PP\null.  Like the $d=3$ ABC system, the
general $d=2$ Lotka--Volterra system has been subjected to Painlev\'e
analyses \mycite{Hua96,Leach2004}.  It appears that the only such systems
with the~PP also have DP's from which a first integral can be constructed;
thus they are at~least partially integrable.  But a Painlev\'e analysis of
general $d=3$ Lotka--Volterra systems has yet to be performed.  And until
now, no explicit solution involving Painlev\'e transcendents of any $d=3$
Lotka--Volterra system seems to have been published.

\subsection{Overview of following results}
\label{subsec:1c}

In \S\,\ref{sec:2} we illustrate our methods by integrating the above
case~II of the $d=2$ Lotka--Volterra system with the aid of a `new
time'~$t$.  The resulting expressions for $x_1,x_2$ and~$\tau$ as functions
of~$t$, involving an incomplete beta function, are new.  This approach is
equivalent to that of \myciteasnoun{Goriely92}, but unlike him we do not
posit a fixed form for the new time transformation $t=t(\tau)$, or make
explicit use of the first integral~(\ref{eq:1stintegral2}).  Equal but
nonzero growth rates are also treated.

In \S\,\ref{sec:3} we obtain our central result: the general solution of
case~$\textrm{II}_i$ of the $d=3$ Lotka--Volterra system, initially
requiring that the growth rates $a_{10},a_{20},a_{30}$ be equal.  Again the
function $t=t(\tau)$ is not constrained \emph{a~priori}.  It turns~out to
satisfy a generalized Schwarzian equation (gSE), which is integrated with
the aid of the incomplete beta function.  No~explicit use is made of the
first integral~(\ref{eq:1stintegral3}).  Many systems are explicitly
solved, including ABC and May--Leonard ones.

Section~\ref{sec:3} relates our solution of case-$\textrm{II}_i$ $d=3$
systems to the results of \myciteasnoun{CartonLeBrun69c}, who classified
all gSE's with the Painlev\'e property.  Her results yield a classification
of equal growth rate case-$\textrm{II}_i$ systems with the property: each
such has elementary or elliptic general solutions $x=x(\tau)$.  In
\S\,\ref{sec:4} we express the general solutions of several $d=3$ systems
with unequal growth rates in~terms of Painlev\'e transcendents.  These may
be the first such solutions ever obtained.

In~\S\,\ref{sec:6} we summarize the results of
\S\S\,\ref{sec:2}--\ref{sec:4}, and make some final remarks.

\section{Two-dimensional integration}
\label{sec:2}

To show how certain Lotka--Volterra systems of the form~(\ref{eq:LV}) can
be integrated parametrically with the aid of a new time variable~$t$,
consider the case\nobreakdash-II $d=2$ systems of table~\ref{tab:2tab}:
ones with $a_{10}=a_{20}=0$, i.e., with zero intrinsic growth rates.  If
the interaction matrix~$\mathsf{A}$ satisfies $a_{11}\neq a_{21}$ and
$a_{22}\neq a_{12}$, by scaling (redefining) the components $x_1,x_2$ one
can scale the columns of~$\mathsf{A}$ so that
\begin{equation}
  \mathsf{A}=\left(
  \begin{array}{cc}
    -a_1 & 1-a_2 \\
    1-a_1 & -a_2
  \end{array}
\right)
\end{equation}
for some $a_1,a_2$.  The resulting rather symmetric system 
\begin{equation}
\label{eq:system2}
\left\{
\begin{aligned}
\dot x_1 &= x_1\left[-a_1\,x_1 + (1-a_2)x_2\right],\\
\dot x_2 &= x_2\left[+(1-a_1)x_1 -a_2\,x_2\right]
\end{aligned}
\right.
\end{equation}has four invariant lines: the usual
$x_1=0$, $x_2=0$ and the one at infinity, which plays no role here; also
$x_1-\nobreak x_2=0$, since if $x_1=x_2$ then $\dot x_1=\dot x_2$.  Only
solutions $x=x(\tau)$ not lying along any of these lines will be
considered.  By examination,
\begin{equation}
\label{eq:gen1st}
  I = \left|x_1\right|^{a_2} \left|x_2\right|^{a_1} \left|x_1-x_2\right|^{1-a_1-a_2}
\end{equation}
is a first integral: the reciprocal of the standard case\nobreakdash-II
first integral~(\ref{eq:1stintegral2}).

Define an auxiliary (new time) variable~$t$ by 
\begin{equation}
\label{eq:tdef}
  t \defeq (x_1 + x_2) / (x_2 - x_1),
\end{equation}
so that $t=-1, 1, \infty$ correspond to the three just-mentioned invariant
lines.  The three $t$\nobreakdash-intervals $(-\infty,-1)$, $(-1,1)$,
$(1,\infty)$ correspond to sectors in the $x_1,x_2$\nobreakdash-plane lying
between the lines.  Thus in the first quadrant, the sectors $0<x_2<x_1$ and
$0<x_1<x_2$ correspond to the $t$\nobreakdash-intervals $(-\infty,-1)$
and~$(1,\infty)$.  To any solution $x=x(\tau)$ of the
system~(\ref{eq:system2}) there is associated a function $t=t(\tau)$,
taking values in one of the three $t$\nobreakdash-intervals.  The
value~$t_0$ taken by~$t$ at the time-origin (say, at $\tau=0$) determines
the $t$\nobreakdash-interval.

By calculus applied to (\ref{eq:tdef}) and the system~(\ref{eq:system2}),
$t=t(\tau)$ satisfies
\begin{equation}
\label{eq:ODE2}
  \frac{\ddot t}{\dot t^2} - \left(
\frac{1-a_1}{t+1} + \frac{1-a_2}{t-1} \right)= 0,
\end{equation}
which integrates to the hyperlogistic growth law
\begin{equation}
\label{eq:tdot}
  \dot t = A\times|t+1|^{1-{a}_1}|t-1|^{1-{a}_2},
\end{equation}
$A\neq0$ being arbitrary.  Hence viewed inversely as a function of~$t$,
$\tau$~is given by
\begin{equation}
\label{eq:introbeta}
\tau = A^{-1} \times \int_{t_0}^t |t'+1|^{{a}_1-1}\,
  |t'-1|^{{a}_2-1}\,dt'.
\end{equation}
Moreover, it follows by differentiating~(\ref{eq:tdef}) and
exploiting~(\ref{eq:system2}) that
\begin{equation}
\label{eq:x1x2}
  (x_1,x_2)=(x_1(t),x_2(t))=\left(\frac{\dot t}{t+1},\,\frac{\dot t}{t-1}\right).
\end{equation}
Thus $x_1,x_2$ as well as~$\tau$ can be expressed as functions of the new
time~$t$.  

The integral in~(\ref{eq:introbeta}) defines an increasing function of~$t$:
an \emph{incomplete beta function} $\tau=\mathrm{B}_{a_1,a_2;t_0}(t)$ with
parameters $a_1,a_2$, as is explained in the appendix.  It maps in an
increasing way the $t$\nobreakdash-interval containing~$t_0$ onto some
$\tau$\nobreakdash-interval, which may be infinite.  The
system~(\ref{eq:system2}) is thus solved parametrically by
\begin{subequations}
\label{eq:parametric2}
\begin{align}
  \tau=\tau(t) &= A^{-1}\times \mathrm{B}_{a_1,a_2;t_0}(t), \label{eq:big1_2}\\
  (x_1,x_2)=(x_1(t),x_2(t))&=A\times|t+1|^{1-{a}_1}|t-1|^{1-{a}_2}\left(\frac1{t+1},\,\frac1{t-1}\right),\label{eq:x1x2t}
\end{align}
\end{subequations}
(\ref{eq:x1x2t}) coming from~(\ref{eq:x1x2}).  The parameter~$t$ is
restricted to the relevant $t$\nobreakdash-interval.

This is a \emph{complete} integration of the $d=2$
system~(\ref{eq:system2}), as the expressions for $x_1,x_2$ and~$\tau$ as
functions of the new time~$t$ involve two undetermined constants:
$A$~and~$t_0$.  (Other than determining the $t$\nobreakdash-interval, the
latter merely shifts~$\tau$.)  Incomplete beta functions are supported by
many software packages, so the numerical solution of any
case\nobreakdash-II $d=2$ system is quite easy.  Growth rates
$a_{10},a_{20}$ that are nonzero but equal can readily be incorporated; see
example~\ref{ex:first} below.

One can also eliminate $t$, and at least formally express $x_1,x_2$ as
functions of the original time~$\tau$.  Let $t={\mathrm
  B}^{-1}_{a_1,a_2}(\tau)$ be any convenient, standardized solution of the
$A=1$ case of the nonlinear ODE~(\ref{eq:tdot}).  (The function ${\mathrm
  B}^{-1}_{a_1,a_2}$ will map in an increasing way some
$\tau$\nobreakdash-interval $(\tau_\textrm{min},\tau_\textrm{max})$, which
may be infinite, onto one of the $t$\nobreakdash-intervals $(-\infty,-1)$,
$(-1,1)$, $(1,\infty)$; so it will depend on the choice of
$t$\nobreakdash-interval.)  The general solution $x=x(\tau)$
of~(\ref{eq:system2}) is then given by (\ref{eq:x1x2}) or~(\ref{eq:x1x2t})
with
\begin{equation}
\label{eq:tdefviaB}
  t={\mathrm B}_{a_1,a_2}^{-1}\bigl(A(\tau-\tau_0)\bigr),
\end{equation}
which is defined if $A(\tau-\tau_0)$ lies
in~$(\tau_\textrm{min},\tau_\textrm{max})$.  The solution $x=x(\tau)$ thus
contains two free parameters, $A$ and~$\tau_0$, and implicitly a choice of
$t$-interval, i.e.\ a choice of sector in the $x_1,x_2$\nobreakdash-plane.

For this $x=x(\tau)$ to be more than formal, an explicit expression for the
(standardized) inverse incomplete beta function $t={\mathrm
  B}_{a_1,a_2}^{-1}(\tau)$ is needed.  As the appendix explains,
expressions are available for some~$a_1,a_2$.  If the set
$\{1/{a}_1,1/{a}_2,\allowbreak 1/(1-\nobreak{a}_1-\nobreak{a}_2)\}$ is any
of $\{1,m,-m\}$ (for $m$ a positive integer), $\{1,\infty,\infty\}$, or
$\{2,2,\infty\}$, the inverse is expressible in~terms of elementary
functions; and if it is any of $\{2,4,4\}$, $\{2,3,6\}$, or $\{3,3,3\}$,
in~terms of elliptic ones.  In each case the inverse extends to a
one-valued function on the complex $\tau$\nobreakdash-plane.  Thus
$x=x(\tau)$ given by (\ref{eq:x1x2t}) and~(\ref{eq:tdefviaB}), containing
fractional powers, is at~most finite-valued.

Table~\ref{tab:3} gives expressions for the (standardized) inverse function
$t={\mathrm B}^{-1}_{a,a}(\tau)$ when $a=0,\frac14,\frac13,\frac12,1$.
Each is one-valued on the $\tau$\nobreakdash-plane, as stated.  It should
also be noted that if $a_1,a_2$ are integers which are positive, or satisfy
$a_1a_2<0$ with $a_1+\nobreak a_2\le0$, the inverse will be an
\emph{algebraic} function (and hence, finite-valued on the
$\tau$\nobreakdash-plane).  In these cases too, $x=x(\tau)$ will be
finite-valued.  In any of the preceding finite-valued cases, the
Lotka--Volterra system~(\ref{eq:system2}) has by~definition a weak form of
the Painlev\'e property.  This result seems to be new.

\begin{example}
\label{ex:first}
  Consider the symmetric system
  \begin{equation}
    \label{eq:14sys}
    \left\{
    \begin{aligned}
      \dot x_1 &= a_{*0}\,x_1 + x_1 (-x_1+3\,x_2)/4, \\
      \dot x_2 &= a_{*0}\,x_2 + x_2 (+3\,x_1-x_2)/4,
    \end{aligned}
    \right.
  \end{equation}
where $x_1,x_2>0$ and $x_1\neq x_2$ for simplicity.  The state variables
$x_1,x_2$ are the populations of species that grow logistically and also
display mutualism: the growth of each is made more rapid by the presence of
the other.  Suppose initially that $a_{*0}=0$, so that this is the case
$a_1=a_2=1/4$ of the system~(\ref{eq:system2}).

The auxiliary variable $t=(x_1+x_2)/(x_2-x_1)$ satisfies the interval
condition $t\in(1,\infty)$ if (initially and hence subsequently)
$0<x_1<x_2$; suppose this to be so.  By table~\ref{tab:3} the standardized
inverse function for this interval is
\begin{equation}
\label{eq:cn}
  t={\mathrm B}^{-1}_{1/4,1/4}(\tau) = \tfrac12[{\rm cn}^2 + {\rm
      cn}^{-2}](\tau/2),
\end{equation}
where ${\rm cn}$ is the Jacobian function with parameter $m=k^2=1/2$.  The
domain $(\tau_\textrm{min},\tau_\textrm{max})$ is $(0,K_{1/4})$, where
$K_{1/4}=\Gamma(\frac14)^2/2\Gamma(\frac12)\approx3.708$.  Substituting
(\ref{eq:cn}) into~(\ref{eq:x1x2}) yields the general $0<x_1<x_2$ solution,
parametrized by $A>0$ and~$\tau_0$:
\begin{equation}
\label{eq:14}
\bigl(x_1(\tau),x_2(\tau)\bigr)=
A\left(
\frac{{\rm sn}^3}{2\,{\rm cn}\,{\rm dn}}
, \,
\frac{2\,{\rm dn}^3}{{\rm sn}\,{\rm cn}}
\right)\bigl( A(\tau-\tau_0)/2\bigr),
\end{equation}
which is defined if $A(\tau-\tau_0)$ lies in~$(0,K_{1/4})$.  Here such
elliptic-function identities as ${\rm sn}^2 + {\rm cn}^2 = 1$ and $m\,{\rm
  sn}^2 + {\rm dn}^2 = 1$ have been used.  For this system the first
integral~(\ref{eq:gen1st}) is $I=|x_1x_2|^{1/4}|x_1-x_2|^{1/2}$, and by
examination $I\equiv\sqrt2\,A$.

The phase portrait in the $0<x_1<x_2$ sector is evident from~(\ref{eq:14}).
The endpoints $\tau_\textrm{min}=0$, $\tau_\textrm{max}=K_{1/4}$, are
zeroes of ${\rm sn},{\rm cn}$ respectively.  So as
$A(\tau-\nobreak\tau_0)\to0^+$, $t\to1^+$ and $(x_1,x_2)$ diverges to
infinity while approaching the positive $x_2$\nobreakdash-axis
(an~invariant line).  Also as $A(\tau-\nobreak\tau_0)\to K_{1/4}^-$,
$t\to\infty$ and $(x_1,x_2)$ diverges to infinity while approaching the
$x_1=x_2$ line (also invariant).  These divergences are due to the strong
mutualism.  Although the solution~(\ref{eq:14}) has a real
$\tau$\nobreakdash-interval of width $A^{-1}K_{1/4}$ as its domain, it
extends in a one-valued way to the complex $\tau$\nobreakdash-plane.  Thus
the system~(\ref{eq:14sys}) has the Painlev\'e property.

Incorporating a nonzero common growth rate~$a_{*0}$ is straightforward, as
noted in the introduction.  Exponentially modifying the independent and
dependent variables yields the general $0<x_1<x_2$ solution, parametrized
by real $\tau_1$ and~$C$:
\begin{equation}
\label{eq:15}
\bigl(x_1(\tau),x_2(\tau)\bigr)=
{\rm e}^{a_{*0}(\tau-\tau_1)}\left(
\frac{{\rm sn}^3}{2\,{\rm cn}\,{\rm dn}}
,\,
\frac{2\,{\rm dn}^3}{{\rm sn}\,{\rm cn}}
\right)\bigl( ({\rm e}^{a_{*0}(\tau-\tau_1)}-C)/2\bigr),
\end{equation}
which is defined if ${\rm e}^{a_{*0}(\tau-\tau_1)}-C$ lies in~$(0,K_{1/4})$.
Compared to the mutualism, the role played by any $a_{*0}>0$ in causing
finite-time blow-up is minor.

On the complex $\tau$\nobreakdash-plane the poles of any instance of the
solution~(\ref{eq:14}) lie on a square lattice, since ${\rm sn},{\rm
  cn},{\rm dn}$ are doubly periodic; but those of any instance
of~(\ref{eq:15}) lie on an exponentially stretched lattice.  The pole
locations are of~interest because solutions in the complex time domain of
non-integrable systems are expected to have irregular patterns of
singularities \mycite{Bessis86}.
\end{example}

The just-concluded example was rather special, as setting $a_1=a_2=1/4$
leads to one-valuedness on the complex $\tau$\nobreakdash-plane.  It must
be stressed that for generic $a_1,a_2$, owing to the presence of branch
points the general complex-domain solution $(x_1(\tau),x_2(\tau))$ of the
system~(\ref{eq:system2}) will not be one-valued or even finite-valued.
The distribution of its poles over its many branches remains to be
explored.

However, the example made clear the behaviour of any real-domain solution
$(x_1(\tau),x_2(\tau))$ in the forward-time limit; i.e., as $\tau$~tends to
the upper endpoint of its interval of definition.  In the first quadrant,
either the invariant line $x_1=x_2$ will be an attractor (as~in the
example), or it will be a repeller and the lines $x_1=0$, $x_2=0$ will be
attractors.  In population ecology, if $a_1,a_2>1$ so that the two species
compete rather than display mutualism, these two possibilities would be
called competitive coexistence and exclusion \mycite{MaynardSmith74}.

\smallskip
The analysis in this section of the case-II $d=2$ system assumed for
simplicity that $a_{11}\neq a_{21}$ and $a_{22}\neq a_{12}$.  This can be
relaxed.  Suppose the growth rates are zero ($a_{10}=a_{20}=0$) and say,
that $a_{11}=a_{21}\eqdef a_{*1}$.  Then by examination the three usual
invariant lines, namely $x_1=0$, $x_2=0$ and the one at infinity, have
respective multiplicities $1,2,1$ in the sense of
\myciteasnoun{Christopher2007}; and there are no others.  (This
configuration of lines is numbered 4.20 by \myciteasnoun{Schlomiuk2010}.)
The line $x_2=0$ thus has an extra (generalized) Darboux polynomial
associated to~it.  The first integral~(\ref{eq:1stintegral2}) is
accordingly replaced by
\begin{equation}
I = \left|x_1\right|^{a_{22}} \left|x_2\right|^{-a_{12}}  \exp(a_{*1}x_1/x_2),
\end{equation}
exhibiting the DP's $x_1,x_2$ and the extra generalized DP\null.  It is
convenient to choose as the new time variable $s\defeq x_1/x_2$.  The
ODE~(\ref{eq:ODE2}) is then replaced by
\begin{equation}
  \frac{\ddot s}{\dot s^2} = \frac1{a_{12}-a_{22}}\left(\frac{a_{12}}s +
  a_{*1}\right).
\end{equation}
This can be integrated to express $\tau=\tau(s)$ in~terms of an incomplete
gamma function, or alternatively (if $a_{22}=-a_{12}$) the error
function~$\textrm{erf}$.  This sheds light on the solution of a particular
Lotka--Volterra system with $a_{11}=a_{21}$, involving $\textrm{erf}$
and~${\textrm{erf}}^{-1}$, that was recently published
\mycite[\S\,4.2]{Chandrasekar2007}.

\section{Three-dimensional integration}
\label{sec:3}

\subsection{General case-$\text{II}_i$ systems}
\label{subsec:3a}

It will now be shown that an unexpectedly large family of $d=3$
Lotka--Volterra systems of the form~(\ref{eq:LV}) can be integrated
parametrically by the technique of~\S\,\ref{sec:2}.  These are
case\nobreakdash-$\textrm{II}_i$ systems as defined in
table~\ref{tab:3tab}; specifically, systems with $a_{ji}=a_{ki}\eqdef
a_{*i}$, so that the species $j,k$ other than~$i$ have equal effects on
species~$i$.  Without loss of generality $i=3$ and $(j,k)=(1,2)$ will
be taken.  Initially it will be assumed that $a_{10}=a_{20}=a_{30}=0$,
i.e., that the intrinsic growth rates are zero, though including rates that
are nonzero but equal is an easy matter.

Systems in this family include certain $\textrm{ABC}$ and symmetric
May--Leonard models, which are treated as examples in
\S\S\,\ref{sec:3}\,$\ref{subsec:3b}$ and~\ref{sec:3}\,$\ref{subsec:3c}$.
Besides constructing explicit solutions, one can determine from powerful
results of \myciteasnoun{CartonLeBrun69c} precisely which systems in the
family have the Painlev\'e property that each solution $x=x(\tau)$ extends
in a one-valued way to the complex $\tau$\nobreakdash-plane.  Generalized
case\nobreakdash-$\textrm{II}_i$ systems which have unequal growth rates
will be examined in \S\,\ref{sec:4}.

Suppose $a_{13}=a_{23}\eqdef a_{*3}$.  If the interaction
matrix~$\mathsf{A}$ satisfies $a_{11}\neq a_{21}$, ${a_{22}\neq a_{12}}$
and $a_{33}\neq a_{*3}\neq0$, by scaling (redefining) the components
$x_1,x_2,x_3$ one can scale the columns of~$\mathsf{A}$ so that
\begin{equation}
  \mathsf{A}=
\left(
\begin{array}{cc|c}
-a_1 & 1-a_2 & 1 \\
1-a_1 & -a_2 & 1 \\
\hline
b_1-a_1 &   b_2-a_2 & -1/n
\end{array}
\right)
\end{equation}
for some $a_1,a_2;b_1,b_2$ and $n\neq0,-1$.  ($n=\infty$ will signify
$1/n=0$.)  The system\begin{equation}
\label{eq:system3}
\left\{
\begin{aligned}
  \dot x_1 &= x_1\left[-a_1x_1 + (1-a_2)x_2 + x_3\right],\\
  \dot x_2 &= x_2\left[+(1-a_1)x_1  -a_2 x_2 + x_3\right],\\
  \dot x_3 &= x_3\left[(b_1-a_1)x_1 + (b_2-a_2)x_2 -
    x_3/n\right],
\end{aligned}
\right.
\end{equation}
which reduces to the $d=2$ system~(\ref{eq:system2}) when $x_3\equiv0$, has
five invariant planes: the usual $x_1=0$, $x_2=0$, $x_3=0$ and the one at
infinity, which plays no role here; also $x_1-\nobreak x_2=0$, since if
$x_1=x_2$ then $\dot x_1=\dot x_2$.  Solutions $x=x(\tau)$ not lying
in any of these planes are of primary interest.  By examination,
\begin{equation}
\label{eq:firstintegral3}
  I =
  \left|x_1\right|^{\bar n a_2-b_2}\left|x_2\right|^{\bar n a_1-b_1}\left|x_3\right|
  \,
  \left|x_1-x_2\right|^{\bar n(1-a_1-a_2)-(1- b_1 - b_2) }
\end{equation}
is the specialization of the case\nobreakdash-$\textrm{II}_3$
first integral~(\ref{eq:1stintegral3}).  Here and below,
\begin{equation}
  \bar n\defeq (n+1)/n
\end{equation}
for convenience; note that $\bar n\neq 0$, with $\bar n=1$ meaning 
$n=\infty$, i.e., $1/n=0$.

Define an auxiliary variable~$t$ as in \S\,\ref{sec:2} by
\begin{equation}
\label{eq:tdef3}
  t \defeq (x_1 + x_2) / (x_2 - x_1),
\end{equation}
so that $t=-1,1,\infty$ correspond to the planes $x_1=0$, $x_2=0$ and
$x_1-\nobreak x_2=0$.  Thus $0<x_2<x_1$ and $0<x_1<x_2$ correspond to the
$t$\nobreakdash-intervals $(-\infty,-1)$ and~$(1,\infty)$.  The function
$t=t(\tau)$ associated to any trajectory $x=x(\tau)$ takes values in one of
the intervals $(-\infty,-1)$, $(-1,1)$, $(1,\infty)$.  On any portion of
$x=x(\tau)$ on which $\dot t$~does not change sign, $t$~can be used as an
alternative parameter: a~new time.

By repeatedly differentiating~(\ref{eq:tdef3}) to obtain formulas for
$t,\dot t, \ddot t$ and $\dddot t$, each time
exploiting~(\ref{eq:system3}), and eliminating $x_1,x_2,x_3$ from these
formulas, one finds
\begin{align}
\label{eq:gSE}
&\frac{\dddot t}{\dot t^3}  + (\bar n - 2)
\left(\frac{\ddot t}{\dot t^2}\right)^2
+\left[
\frac{1-b_1-2\bar n(1-a_1)}{t+1} + \frac{1-b_2-2\bar n(1-a_2)}{t-1}
\right]\frac{\ddot t}{\dot t^2}\nonumber\\
&\quad{}+\biggl\{
\frac{(1-a_1) \left[b_1+\bar n(1-a_1)\right]}{(t+1)^2} +
\frac{(1-a_2) \left[b_2+\bar n(1-a_2)\right]}{(t-1)^2}\\
&\qquad\quad{}-\tfrac{{(1-a_1) \left[b_1+\bar n(1-a_1)\right]}+{(1-a_2) \left[b_2+\bar n(1-a_2)\right]}+(2-a_1-a_2)\left[(1-b_1-b_2)-\bar n(2-a_1-a_2)\right]}{(t+1)(t-1)}
\biggr\}
= 0,\nonumber
\end{align}
which is the $d=3$ counterpart of~(\ref{eq:ODE2}).  The nonlinear
third-order ODE~(\ref{eq:gSE}) will be called a \emph{generalized
  Schwarzian equation} (gSE), since when $\bar n=1/2$ (i.e., $n=-2$) its
first two terms constitute a Schwarzian derivative, and if additionally
$b_1=a_1$, $b_2=a_2$ so that its two $\ddot t/\dot t^2$ terms vanish, it
becomes a Schwarzian equation of a type familiar from the conformal mapping
of triangles \mycite{Nehari52}.

The seemingly complicated gSE~(\ref{eq:gSE}), with solution $t=t(\tau)$,
can be integrated without difficulty.  Suppose for simplicity that $\dot
t>0$.  Let $\dot t\eqdef f^{-1/\bar n}$, i.e., $f\defeq \dot t^{\bar n}$,
and view $t,f$ (rather than $\tau,t$) as the independent and dependent
variables.  By applying the chain rule ${\rm d}/{\rm d}\tau=\dot t\,D_t =
f^{1/\bar n}D_t$ where $D_t\defeq {\rm d}/{\rm d}t$, one finds that
\begin{align}
&\Biggl\{D_t^2 + 
\biggl[
\frac{1-b_1-2\bar n(1-a_1)}{t+1} + \frac{1-b_2-2\bar n(1-a_2)}{t-1}
\biggr] D_t\nonumber\\
&\ {}+\bar n
\biggl\{
\frac{(1-a_1) \left[b_1+\bar n(1-a_1)\right]}{(t+1)^2} +
\frac{(1-a_2) \left[b_2+\bar n(1-a_2)\right]}{(t-1)^2}\label{eq:PE}\\
&\qquad{}-\tfrac{{(1-a_1) \left[b_1+\bar n(1-a_1)\right]}+{(1-a_2) \left[b_2+\bar n(1-a_2)\right]}+(2-a_1-a_2)\left[(1-b_1-b_2)-\bar n(2-a_1-a_2)\right]}{(t+1)(t-1)}
\biggr\}
\Biggr\}\,f=0    \nonumber
\end{align}
is the ODE (surprisingly, a linear one) satisfied by $f=f(t)$.  This ODE is
a so-called Papperitz equation, which can be greatly simplified by a
substitution of a standard type.  Substituting $f=(t+1)^{\bar
  n(1-a_1)}(t-1)^{\bar n(1-a_2)}\,g$ reduces it to the degenerate
hypergeometric equation
\begin{equation}
\label{eq:degenhyperg}
\left\{D_t^2 + \left[\frac{1-b_1}{t+1}+ \frac{1-b_2}{t-1}\right] D_t\right\}\,g=0,
\end{equation}
which can be integrated by inspection.  In this way one deduces that
\begin{align}
  \dot t^{\bar n} = f(t) &= \left|t+1\right|^{\bar n(1-a_1)}\left|t-1\right|^{\bar n(1-a_2)} \left[K_1 + K_2 \int_{t_0}^t | t'+1 |^{b_1-1}  |t'-1|^{b_2-1} \,dt'\right]\nonumber\\
 &= \left|t+1\right|^{\bar n(1-a_1)}\left|t-1\right|^{\bar n(1-a_2)} \left[K_1 + K_2\,{\rm B}_{b_1,b_2;t_0}(t)\right],\label{eq:betterone}
\end{align}
where $K_1,K_2$ are undetermined constants and $t_0$~is any convenient
point in the relevant $t$\nobreakdash-interval.  For the incomplete beta
function ${\rm B}_{b_1,b_2;t_0}(t)$, see the appendix.

Formulas for the inverse function $\tau=\tau(t)$ and the alternatively
parametrized trajectory $x=x(t)$ of the system~(\ref{eq:system3}) follow
immediately from~(\ref{eq:betterone}).  The restriction $\dot t>0$ will now
be dropped: suppose instead that on the portion of the trajectory under
study, $\pm\dot t>0$.  By integrating ${\rm d}\tau/{\rm d}t=\dot t^{-1}$
one obtains
\begin{subequations}
\label{eq:parametric3}
\begin{gather}
\begin{aligned}
\tau = \tau(t) &= \tau_0 \pm \int_{t_0}^t   
|t'+1|^{a_1-1}|t'-1|^{a_2-1}
\left|K_1 + K_2\,{\rm B}_{b_1,b_2;t_0}(t')\right|^{1/(n+1)-1}\,dt',
\end{aligned}\label{eq:big1}\\[-1.25pt]
\begin{aligned}
(x_1,x_2,x_3) = \left(x_1(t),x_2(t),x_3(t)\right) &=  \left(\tfrac{\dot t}{t+1},\tfrac{\dot t}{t-1},\tfrac{\ddot t}{\dot t} - (1-a_1)\tfrac{\dot t}{t+1} - (1-a_2)\tfrac{\dot t}{t-1}\right).\label{eq:big2}
\end{aligned}
\end{gather}
\end{subequations}
In (\ref{eq:big1}), $t$~is restricted to the relevant
$t$\nobreakdash-interval (containing any conveniently chosen point~$t_0$)
and is further restricted by the requirement that the quantity $K_1
+\nobreak K_2\,{\rm B}_{b_1,b_2;t_0}(t')$, within absolute value signs, not
change sign.  The accompanying formulas~(\ref{eq:big2}), which extend the
$d=2$ formula~(\ref{eq:x1x2}), are an easy exercise.  They come by
reverting the abovementioned expressions for $t,\dot t,\ddot t$ as rational
functions of~$x_1,x_2,x_3$.  In practice one would compute $\dot t,\ddot t$
in~(\ref{eq:big2}) by applying $\dot t=({\rm d}\tau/{\rm d}t)^{-1}$ and
$\ddot t=\dot t({\rm d}/{\rm d}t)\dot t$ to~(\ref{eq:big1}).

The parametric solution $(\tau,x)=(\tau(t),x(t))$ displayed
in~(\ref{eq:parametric3}) is the central result of this paper.  It is a
\emph{complete} integration of the $d=3$ Lotka--Volterra
system~(\ref{eq:system3}), as the expressions for $\tau$ and $x_1,x_2,x_3$
as functions of~$t$ involve three undetermined constants: the coefficients
$K_1,K_2$, and also~$\tau_0$ (which merely shifts~$\tau$).  Growth rates
$a_{10},a_{20},a_{30}$ that are nonzero but equal can easily be
incorporated in~(\ref{eq:parametric3}), as when $d=2$.  The solution
(\ref{eq:parametric3}) subsumes the $d=2$ solution~(\ref{eq:parametric2}).
This is because if $K_2=0$ then $\tau=\tau(t)$ given by~(\ref{eq:big1})
reduces to~(\ref{eq:big1_2}), and $x_3=x_3(\tau)$ computed
from~(\ref{eq:big2}) will be identically zero.

\begin{example}
\label{ex:31}
Consider a $d=3$ Lotka--Volterra system of the form~(\ref{eq:system3}) with
$b_1=a_1$, $b_2=a_2$, so that species~3 is unaffected by species 1 and~2.
In this case $(\tau,x)=(\tau(t),x(t))$ can be simplified, since if
$K_2\neq0$, Eq.~(\ref{eq:big1}) integrates to
\begin{equation}
\tau=\tau(t) = \tau_0\pm
\begin{cases}
|K_2|^{-1}\bigl\{|K_1+K_2\,{\rm B}_{a_1,a_2;t_0}(t)|^{1/(n+1)} - |K_1|^{1/(n+1)}\bigr\} & n\textrm{ finite},\\
|K_2|^{-1}\ln |1+(K_2/K_1){\rm B}_{a_1,a_2;t_0}(t)| & n=\infty.
\end{cases}
\end{equation}
In~terms of a standardized inverse beta function $t={\rm
  B}^{-1}_{a_1,a_2}(\tau)$ as discussed in the appendix (and used
in~\S\,\ref{sec:2}), this simply says that $t=t(\tau)$ is of the form
\begin{equation}
\label{eq:star}
  t=t(\tau)=
\begin{cases}
{\rm B}^{-1}_{a_1,a_2}\left((A\tau+B)^{n+1}+C\right), & n\text{ finite},\\
 {\rm B}^{-1}_{a_1,a_2}\left(\exp(A\tau+B)+C\right), & n=\infty,
\end{cases}
\end{equation}
for some $A\neq0,B,C$.  For $a_1,a_2$ for which ${\rm
  B}^{-1}_{a_1,a_2}(\tau)$ is expressible in~terms of standard functions,
such as in those in table~\ref{tab:3}, this allows $x=x(\tau)$ when
$b_1=a_1$, $b_2=a_2$ to be similarly expressed, by substituting
(\ref{eq:star}) into~(\ref{eq:big2}).

For many~$a_1,a_2$, such as those in the table, the function $t={\rm
  B}^{-1}_{a_1,a_2}(\tau)$ has a one-valued extension to the complex
$\tau$\nobreakdash-plane.  For such $a_1,a_2$, if $n$~is an \emph{integer}
($n\neq0,-1$) or $n=\infty$, the system~(\ref{eq:system3}) with $b_1=a_1$,
$b_2=a_2$ will have the Painlev\'e property.  This is because if $n$~is an
integer or~$\infty$, $t=t(\tau)$ in~(\ref{eq:star}) and likewise
$x=x(\tau)$ will extend to the $\tau$\nobreakdash-plane with no branchpoint
at $\tau=-B/A$, i.e., in a one-valued way.  But in each such system, the
third species decouples.
\end{example}

One can also determine which $d=3$ Lotka--Volterra systems of the
form~(\ref{eq:system3}), \emph{with no species decoupled}, have the
Painlev\'e property; and in~fact, integrate each such explicitly.  In the
gSE~(\ref{eq:gSE}) for $t=t(\tau)$ there are seven terms and thus seven
coefficients, the first two being $1$ and $\bar n-\nobreak 2 =\allowbreak
1/n-\nobreak 1$.  Suppose the remaining five are not determined by
$a_1,a_2;b_1,b_2$ and~$n$, as here, but are free to vary.  For some choices
of these five the resulting nonlinear ODE will have the Painlev\'e
property.  By a mathematically rigorous analysis
\myciteasnoun{CartonLeBrun69c} determined, classified and tabulated the
many possibilities, and integrated each resulting ODE\null.  Hence all one
needs to do is compare the coefficients in the gSE~(\ref{eq:gSE}) against
her tables to determine for which $a_1,a_2;b_1,b_2$ and~$n$ all solutions
$t=t(\tau)$ of the gSE will extend in a one-valued way to the
$\tau$\nobreakdash-plane.  For each such choice one can compute $x=x(\tau)$
from~(\ref{eq:big2}), and it will extend similarly; thus the
case\nobreakdash-$\textrm{II}_3$ system~(\ref{eq:system3}) as~well as the
gSE will have the Painlev\'e property.

It is not hard to see that for the gSE~(\ref{eq:gSE}) to have the
Painlev\'e property the parameter~$n$ must be an integer ($n\neq0,-1$ as
usual) or~$\infty$; else the gSE would have complex-domain solutions
$t=t(\tau)$ with movable branchpoints.  The tables of Carton-LeBrun include
infinite families of gSE's in which $n$~is otherwise unconstrained, such as
ones with $b_1=a_1$, $b_2=a_2$ that were implicitly encountered in
example~\ref{ex:31}.  They also include many gSE's with $n=1,2,3,5$
or~$\infty$.  Each can be twice integrated to yield a polynomial identity
$P(t,\dot t;K_1,K_2)=0$ with parameters $K_1,K_2$, the identities appearing
in her table~VIII\null.  Any solution $t=t(\tau)$ is thus reduced to
quadratures.  For nearly all of these gSE's, solutions $t=t(\tau)$ and
hence trajectories $x=x(\tau)$ can be expressed in~terms of elliptic
functions (lemniscatic or equianharmonic); for a few, in~terms of
elementary functions.

\begin{example}
  Consider the $d=3$ Lotka--Volterra system
  \begin{equation}
    \label{eq:sys1eg}
    \left\{
    \begin{aligned}
      \dot x_1 &= a_{*0}\,x_1 + x_1 (x_2+x_3), \\
      \dot x_2 &= a_{*0}\,x_2 + x_2 (x_1+x_3), \\
      \dot x_3 &= a_{*0}\,x_3 + x_3 (x_1-2\,x_2-x_3),
    \end{aligned}
    \right.
  \end{equation}
where for simplicity $x_1,x_2,x_3>0$ and $x_1\neq x_2$.  The state
variables $x_1,x_2,x_3$ are the populations of three species with common
intrinsic growth rate~$a_{*0}$.  If $a_{*0}>0$ then in isolation species~3,
though not 1~or~2, grows logistically.  Species 1 and~2 display mutualism,
as do 1 and~3; and species~2 preys on species~3.

Since a nonzero~$a_{*0}$ is easy to incorporate, set $a_{*0}=0$.
System~(\ref{eq:sys1eg}) is then the case $(a_1,a_2;\allowbreak b_1,b_2;n)
=\allowbreak (0,0;\allowbreak 1,-2;1)$ of~(\ref{eq:system3}).  With these
values the gSE~(\ref{eq:gSE}) for $t=t(\tau)$ is a variant of
Carton-LeBrun's $n=1$ case~II\null.  Integrating twice yields
\begin{equation}
\label{eq:hasrightside}
   \dot t^2 = K_1(t^2-1)^2 + K_2(t+1)^2,
\end{equation}
where $K_1,K_2$ are undetermined constants.  (As $\dot t^{\bar n}=\dot
t^2=f(t)$, the right side is the general solution of the Papperitz
equation~(\ref{eq:PE}).)  Integrating once more yields
\begin{equation}
\label{eq:tsinh}
  t= (x_1+x_2)/(x_2-x_1) = \frac{3+(c_1-c_2\,\cos\tau)}{1-(c_1-c_2\,\cos\tau)}
\end{equation}
as the general solution of the gSE\null.  Here
$c_1,c_2$ are constrained by $c_1^2-\nobreak c_2^2=1$ and $\tau$~can be
replaced by $A\tau+\nobreak B$, for any $A\neq0$ and~$B$.

Substituting (\ref{eq:tsinh}) into~(\ref{eq:big2}) yields
\begin{multline}
  (x_1,x_2,x_3)(\tau) =\\
\left(
\frac{c_2\,\sin\tau}{1-(c_1-c_2\cos\tau)},\,
\frac{2\,c_2\,\sin\tau}{1-(c_1-c_2\cos\tau)^2},\,
\frac{-c_2+(1+c_1)\cos\tau}{[1+(c_1-c_2\cos\tau)]\sin\tau}
\right)
\label{eq:sys1egsoln}
\end{multline}
as the general solution $x=x(\tau)$ of the system~(\ref{eq:sys1eg}) with
$a_{*0}=0$, it being understood that $\tau$~can be replaced by
$A\tau+\nobreak B$, with $x$~scaled by~$A$.  There are three degrees of
freedom: $A$, $B$ and the point on the hyperbola $c_1^2-\nobreak c_2^2=1$.

The first integral~$I$ of~(\ref{eq:firstintegral3}) specializes to
$x_1^2|x_2^{-1}x_3|$, and by examination $I\equiv A^2 \times\nobreak
|c_1+1|/2$.  The existence of a explicit general solution facilitates the
study of global dynamics.
\end{example}

\begin{example}
  Consider the $d=3$ Lotka--Volterra system
  \begin{equation}
    \label{eq:sys2eg}
    \left\{
    \begin{aligned}
      \dot x_1 &= x_1 \left[-\tfrac1r\,x_1 + (1+\tfrac1r)x_2 + x_3\right], \\
      \dot x_2 &= x_2 \left[+(1-\tfrac1r)x_1 + \tfrac1r\,x_2 + x_3\right], \\
      \dot x_3 &= x_3 \left[\tfrac3r\,x_1-\tfrac3r\,x_2 - x_3\right],
    \end{aligned}
    \right.
  \end{equation}
where $r\neq0$, and for simplicity $x_1,x_2,x_3>0$ and $x_1\neq x_2$.  The
system~(\ref{eq:sys1eg}) is the case $(a_1,a_2;\allowbreak b_1,b_2;n)
=\allowbreak (\frac1r,-\frac1r;\allowbreak \frac4r,-\frac4r;1)$
of~(\ref{eq:system3}).  With these values the gSE~(\ref{eq:gSE}) for
$t=t(\tau)$ is a variant of Carton-LeBrun's case~e.IV\null.  The gSE can be
integrated twice to yield $t=\allowbreak (u^r+\nobreak 1)/\allowbreak
(u^r-\nobreak 1)$, where $u$~is a solution of
\begin{equation}
\label{eq:starstar}
  \dot u^2 = K_1 u^4 - K_2
\end{equation}
and $K_1,K_2$ are undetermined constants.  (To confirm this, it is easiest
not to integrate the gSE itself, but rather the Papperitz
equation~(\ref{eq:PE}) or preferably the degenerate hypergeometric
equation~(\ref{eq:degenhyperg}), which are equivalent.)  By
integrating~(\ref{eq:starstar}) once more, one deduces that $u\propto{\rm
  cn}(\tau)$ and thus that
\begin{equation}
  \label{eq:t2}
  t=(x_1+x_2)/(x_2-x_1) = \frac{C\,{\rm cn}^r + 1}{C\,{\rm cn}^r - 1}(\tau).
\end{equation}
Here $\rm cn$ is the `lemniscatic' Jacobian elliptic function with modular
parameter $m=k^2=1/2$, and $C$~is arbitrary; and as above, $\tau$~can be
replaced by $A\tau+\nobreak B$.

Substituting the expression (\ref{eq:t2}) 
into~(\ref{eq:big2}) yields
\begin{equation}
  (x_1,x_2,x_3)(\tau) =
\left(
\frac{r\,{\rm sn}\,{\rm dn}}{(C\,{\rm cn}^r-1){\rm cn}},\,
\frac{C\,r\,{\rm sn}\,{\rm dn}\,{\rm cn}^{r-1}}{C\,{\rm cn}^r-1},\,
\frac{{\rm cn}^3}{{\rm sn}\,{\rm dn}}
\right)(\tau)
\label{eq:sys2egsoln}
\end{equation}
as the general solution $x=x(\tau)$ of the system~(\ref{eq:sys2eg}), it
being understood that $\tau$~can be replaced by $A\tau+\nobreak B$, with
$x$~scaled by~$A$.  The first integral~$I$ of~(\ref{eq:firstintegral3})
specializes to $|x_1/x_2|^{2/r}|x_3(x_1-x_2)|$, and by examination
$I\equiv\allowbreak A^2 \times |r||C|^{-2/r}$.  If $r$~is an integer then
(\ref{eq:sys2egsoln}) like~(\ref{eq:sys1egsoln}) will be one-valued on the
complex $\tau$\nobreakdash-plane; thus the system (\ref{eq:sys2eg})
like~(\ref{eq:sys1eg}) will have the Painlev\'e property.
\end{example}

\subsection{$\textrm{ABC}$ systems}
\label{subsec:3b}

An interesting application of the results of the last subsection is to a
well-known family of ABC systems, which as explained
in~\S\,\ref{sec:1}\,$\ref{subsec:1b}$ are (generic) $d=3$ Lotka--Volterra
systems without self-interactions.  Any ABC system, with interaction matrix
$\mathsf{A}=(a_{ij})_{i,j=1}^3$ given by~(\ref{eq:AABC}), is of the form
\begin{equation}
\label{eq:ABCsystem}
\left\{
  \begin{aligned}
    \dot y_1 &= y_1(A_2\,y_2 + y_3),\\
    \dot y_2 &= y_2(A_3\,y_3 + y_1),\\
    \dot y_3 &= y_3(A_1\,y_1 + y_2).
  \end{aligned}
\right.
\end{equation}
It is known that when one of $A_1,A_2,A_3$ equals unity the ABC system is
completely integrable in that a pair of first integrals exists, one of them
being of the Darboux polynomial type \mycite{Grammaticos90}.  But no
general solution of ABC systems of this type has previously been published.
From our point of view the integrability comes from an ABC system with
$A_i=1$ being a case\nobreakdash-${\rm II}_i$ system as defined in
table~\ref{tab:3tab}.  This facilitates a complete integration.

Suppose without loss of generality that $A_3=1$, and also that $A_2\neq0$.
In~terms of a scaled state vector $(x_1,x_2,x_3)\defeq (y_1,A_2y_2,y_3)$
the system becomes
\begin{equation}
\label{eq:normalizedABCsystem}
\left\{
  \begin{aligned}
    \dot x_1 &= x_1(x_2 + x_3),\\
    \dot x_2 &= x_2(x_1 + x_3),\\
    \dot x_3 &= x_3(A_1\,x_1 + A_2^{-1}x_2),
  \end{aligned}
\right.
\end{equation}
which is the $(a_1,a_2;b_1,b_2;n)=(0,0;A_1,A_2^{-1};\infty)$ case of the
case\nobreakdash-${\rm II}_3$ system~(\ref{eq:system3}).  An integration
of~(\ref{eq:normalizedABCsystem}) therefore follows at~once from the
complete integration~(\ref{eq:parametric3}).  (As~above, solutions
that do not lie in any of the invariant planes $x_1=0$, $x_2=0$,
$x_1-\nobreak x_2=0$ are of primary interest.)

The new time $t\defeq (x_1+x_2)/(x_2-x_1)$ is used as the parameter.  Along
any segment of a trajectory $x=x(\tau)$ on which $\pm (\dot t={\rm d}t/{\rm
  d}\tau)>0$, the variables $\tau$~and $x=(x_1,x_2,x_3)$ are expressed as
functions of~$t$ by
\begin{subequations}
\label{eq:ABCsoln}
\begin{gather}
\begin{aligned}
\tau(t) &= \tau_0 \pm \int_{t_0}^t   
\left|\,\bigl[(t')^2-1\bigr]\,\bigl[K_1 + K_2\,{\rm B}_{A_1,A_2^{-1};\,t_0}(t')\bigr]\,\right|^{-1}dt',
\end{aligned}\label{eq:big1new}\\
\begin{aligned}
\left(x_1(t),x_2(t),x_3(t)\right) &=  \left(\frac{\dot t}{t+1},\,\frac{\dot t}{t-1},\,\frac{\ddot t}{\dot t} - \frac{\dot t}{t+1} - \frac{\dot t}{t-1}\right).\label{eq:big2new}
\end{aligned}
\end{gather}
\end{subequations}
Here $\tau_0$ is the initial time, at which $t=t_0$.  The parameter~$t$ is
restricted to whichever of $(-\infty,-1)$, $(-1,1)$, $(1,\infty)$
contains~$t_0$.  It may be further restricted by the additional condition
that $K_1 +\nobreak K_2\,{\rm B}_{A_1,A_2^{-1};\,t_0}(t')$, within the
absolute value signs, not change sign over the integration interval.  The
derivatives $\dot t,\ddot t$ in~(\ref{eq:big2new}) are computed by applying
$\dot t=({\rm d}\tau/{\rm d}t)^{-1}$ and $\ddot t=\dot t({\rm d}/{\rm
  d}t)\dot t$ to~(\ref{eq:big1new}).

The parametric general solution~(\ref{eq:ABCsoln}) is a complete
integration of the ABC system with $A_3=1$, since the expressions for
$\tau$ and $x_1,x_2,x_3$ involve three undetermined constants: $K_1,K_2$
and also $\tau_0$ (which merely shifts~$\tau$).  For many choices of
$A_1,A_2$ the expressions for $\tau$ and $x_1,x_2,x_3$ in~terms of~$t$ can
be expressed using elementary functions.  This is because for many choices
of~$b_1,b_2$ the incomplete beta function ${\rm B}_{b_1,b_2;t_0}(t)$
in~(\ref{eq:big1new}) can be so expressed.  The evaluation of $\tau$ and
$x_1,x_2,x_3$ is especially easy if ${\rm B}_{b_1,b_2;t_0}(t)$ is a
rational function of~$t$.

\begin{example}
Consider the case $(A_1,A_2,A_3)=(1,1,1)$ of the ABC
system~(\ref{eq:ABCsystem}) and hence of~(\ref{eq:normalizedABCsystem}),
for which $(a_1,a_2;b_1,b_2;n)=\allowbreak (0,0;1,1;\infty)$.  When $t>1$,
as is the case for any trajectory $x=x(\tau)$ in the sector $0<x_1<x_2$,
the definition~(\ref{eq:Bdef}) implies that ${\rm B}_{1,1;t_0}(t)$ equals
$t$~plus a constant.  Substituting into~(\ref{eq:ABCsoln}) and evaluating
the integral and derivatives yields
\begin{subequations}
\label{eq:gensolneg1}
\begin{align}
\tau(t) &= \frac{\ln(t+C)}{C^2-1}
+\frac{\ln(t+1)}{2(1-C)}  +\frac{\ln(t-1)}{2(1+C)},\label{eq:ABCegtau}\\
(x_1,x_2,x_3)(t) &= \left(
(t-1)(t+C),\, (t+1)(t+C),\, t^2-1
\right),
\end{align}
\end{subequations}where $C$~is a constant; and one can also replace $\tau$ by $A\tau+\nobreak
B$ and scale $(x_1,x_2,x_3)$ by~$A$.  Thus there are three undetermined
constants in~all.

The solution $x=x(\tau)$ in the sector is generated parametrically by
allowing $t$ to range over $(1,\infty)$, or perhaps (depending on~$C$) a
proper sub-interval.  (The latter is because $t>1$ covers $0>x_1>x_2$
as~well as $0<x_1<x_2$.)  The first integral~$I$
of~(\ref{eq:firstintegral3}) specializes to
$|x_1^{-1}x_2^{-1}x_3(x_1-x_2)^{2}|$, and by examination $I\equiv 4|A|$.

Trajectories in other sectors, such as $0<x_2<x_1$, are generated
similarly.  The general solution~(\ref{eq:gensolneg1}) resembles but is
more explicit than a solution of the ABC system with
$(A_1,A_2,A_3)=(1,1,1)$ obtained by \myciteasnoun[\S\,7]{Abenda2001}.
\end{example}


For any ABC system of the form~(\ref{eq:normalizedABCsystem}), the new time
$t=t(\tau)$ satisfies the generalized Schwarzian equation~(\ref{eq:gSE})
with $(a_1,a_2;b_1,b_2;n)=\allowbreak (0,0;A_1,A_2^{-1};\infty)$.  (That
$n=\infty$ implies $\bar n=1$.)  Several of the gSE's listed by
\myciteasnoun{CartonLeBrun69c} as having the Painlev\'e property, in that
all solutions $t=t(\tau)$ extend to a one-valued way to the complex plane,
have $n=\infty$ and $a_1=a_2=0$.  They therefore give rise to ABC systems
with $A_3=1$ that have the Painlev\'e property.

\begin{example}
Consider the case $(A_1,A_2,A_3)=(-1/2,-2,1)$ of the ABC
system~(\ref{eq:ABCsystem}) and hence of~(\ref{eq:normalizedABCsystem}),
for which $(a_1,a_2;b_1,b_2;n)=\allowbreak (0,0;-1/2,-1/2;\infty)$.  With
these values the gSE~(\ref{eq:gSE}) satisfied by $t=t(\tau)$ is a variant
of Carton-LeBrun's case~e.IV.3.  The solution of this gSE is
\begin{equation}
\label{eq:ttanh}
  t(\tau) = \left[(C\tanh\tau) + (C\tanh\tau)^{-1}\right]/\,2,
\end{equation}
where $C\neq0$ is arbitrary and $\tau$~can be replaced by $A\tau+B$.  (This
is best shown by integrating the hypergeometric
equation~(\ref{eq:degenhyperg}), though one could also integrate
Eq.~(\ref{eq:big1new}), as in the last example.)  Substituting
(\ref{eq:ttanh}) into~(\ref{eq:big2new}) gives
\begin{align}
\label{eq:ABCspecialint}
&(x_1,x_2,x_3)(\tau)=\\
&\left(
\frac{-{\rm cosh}+C\,{\rm sinh}}{{\rm sinh}\,{\rm cosh}\,({\rm cosh}+C\,{\rm sinh})}
,\,
\frac{-{\rm cosh}-C\,{\rm sinh}}{{\rm sinh}\,{\rm cosh}\,({\rm cosh}-C\,{\rm sinh})}
,\,
\frac{2(1-C^2)\,{\rm sinh}\,{\rm cosh}}{C^2\,{\rm sinh}^2-{\rm cosh}^2}
\right)(\tau),\nonumber
\end{align}
where again $\tau$~can be replaced by $A\tau+B$, in which case
$(x_1,x_2,x_3)$ is scaled by~$A$.  Thus there are three undetermined
constants in~all.  

Equation~(\ref{eq:ABCspecialint}) provides a complete integration of the
ABC system with $(A_1,A_2,A_3)=(-1/2,-2,1)$.  If $-1<C<1$ the trajectory
defined by~(\ref{eq:ABCspecialint}) with $\tau<0$ lies in the positive
orthant; and if $C=0$ it lies in the plane $x_1-\nobreak x_2=0$.  If
$C=\pm1$ it lies in the plane $x_3=0$.  The first integral~$I$ of
(\ref{eq:firstintegral3}) specializes to $|x_1x_2|^{1/2}|x_3/(x_1-x_2)|$,
and by examination $I\equiv |A|\times |(C^2-\nobreak 1)/2C|$.

This system is of much interest, as it is the $A_3=1$ member of a family of
ABC systems (with $A_1=\allowbreak -1/(A_3+\nobreak1)$ and $A_2=\allowbreak
-(A_3+\nobreak1)/A_3$ for $A_3\neq\allowbreak 0,-1$), each of which has the
Painlev\'e property and is completely integrable \mycite{Bountis84}.  Each
has a first integral that is quadratic in $x_1,x_2,x_3$
\mycite{Strelcyn88}.  If $A_3=1$ this extra first integral is
$J=(x_1-\nobreak x_2)^2+\allowbreak 4x_3(x_1+\nobreak x_2+\nobreak x_3)$,
and by examination $J\equiv A^2\times 16$, irrespective of the integration
constant~$C$.

Bountis et~al.\ remark that the member with $(A_1,A_2,A_3)=(-1/2,-2,1)$ can
be integrated with the aid of elliptic functions; but as
Eq.~(\ref{eq:ABCspecialint}) reveals, $\sinh$ and $\cosh$ suffice.  This
explicit general solution also makes manifest the one-valuedness of the
continuation of $x=x(\tau)$ to the complex $\tau$\nobreakdash-plane.
\end{example}

\subsection{Symmetric May--Leonard systems}
\label{subsec:3c}

A particularly striking application of the results
of~\S\,\ref{sec:3}\,$\ref{subsec:3a}$ is to the integration of certain
May--Leonard systems.  As was explained
in~\S\,\ref{sec:1}\,$\ref{subsec:1b}$, the three species in any
May--Leonard system compete cyclically, with $\alpha,\beta$ being the
strength of clockwise and counter-clockwise interactions.  The systems of
interest here are those with $\alpha=\beta$, for which the species are
equivalent; they are completely integrable.  An explicit general solution
will be obtained for the first time: see
Eqs.\ (\ref{eq:MLode})--(\ref{eq:MLx}) below.  The system with
$\alpha=\beta=-1$, which is mutualistic rather than competitive, will be
shown to have the Painlev\'e property.  May--Leonard systems have been
analysed from the Painlev\'e point of view \mycite{Bountis82,Sachdev97} and
multiple first integrals have been discovered
\mycite{Llibre2011,Tudoran2012}.  But the results below are much more
explicit.

Suppose that $a_{10}=a_{20}=a_{30}=0$, i.e., that the intrinsic growth
rates are zero, since including rates that are nonzero but equal is easy.
The system~(\ref{eq:LV}), with the May--Leonard interaction
matrix~(\ref{eq:MLmatrix}) and $\alpha=\beta$, is then of the form
\begin{equation}
  \left\{
  \begin{aligned}
    \dot y_1 &= y_1(-y_1-\alpha\, y_2 -\alpha\, y_3), \\
    \dot y_2 &= y_2(-\alpha\, y_1-y_2 -\alpha\, y_3), \\
    \dot y_3 &= y_3(-\alpha\, y_1-\alpha\, y_2 - y_3). \\
  \end{aligned}
  \right.
\end{equation}
Suppose that $\alpha\neq0,1$.  Then in~terms of a scaled state vector
$(x_1,x_2,x_3)\defeq\allowbreak \left( (1-\nobreak\alpha)y_1,\,\allowbreak
(1-\nobreak\alpha)y_2,\,-\alpha y_3\right)$, the system becomes
\begin{equation}
  \left\{
  \begin{aligned}
    \dot x_1 &= x_1(-\tfrac{1}{1-\alpha}\,x_1-\tfrac{\alpha}{1-\alpha}\, x_2 +x_3), \\
    \dot x_2 &= x_2(-\tfrac{\alpha}{1-\alpha}\,x_1-\tfrac{1}{1-\alpha}\, x_2 +x_3), \\
    \dot x_3 &= x_3(-\tfrac{\alpha}{1-\alpha}\,x_1-\tfrac{\alpha}{1-\alpha}\, x_2 +\tfrac{1}{\alpha}\, x_3), \\
  \end{aligned}
  \right.
\end{equation}
which is the $(a_1,a_2;b_1,b_2;n)=\allowbreak \left(
(1-\nobreak\alpha)^{-1}\!,\,(1-\nobreak\alpha)^{-1};\,1,1;\,\allowbreak
-\alpha\right)$ case of the case\nobreakdash-${\rm II}_3$
system~(\ref{eq:system3}).  An integration scheme therefore follows at~once
from the complete integration~(\ref{eq:parametric3}).  As~above, it will
yield all solutions that do \emph{not} lie in any of the invariant planes
$x_1=0$, $x_2=0$, $x_1-\nobreak x_2=0$.

The new time variable $t\defeq (x_1+x_2)/(x_2-x_1)$ plays its familiar
role.  Along any segment of a trajectory $x=x(\tau)$ on which $\pm (\dot
t={\rm d}t/{\rm d}\tau)>0$, the function $t=t(\tau)$ satisfies
\begin{equation}
\label{eq:MLode}
\dot t = \pm\left|(t^2-1)(K_1\, t+ K_2)\right|^{\alpha/(\alpha-1)}
\end{equation}
where $K_1,K_2$ are constants.  This is a specialization
of~(\ref{eq:big1}), since according to the definition~(\ref{eq:Bdef}),
${\rm B}_{1,1;t_0}(t)$ equals $t$~plus a constant.  After obtaining
$t=t(\tau)$ by integrating Eq.~(\ref{eq:MLode}), one would substitute it
into~(\ref{eq:big2}), i.e., into
\begin{equation}
\label{eq:MLx}
\left(x_1(\tau),x_2(\tau),x_3(\tau)\right) =  \left(\frac{\dot t}{t+1},\,\frac{\dot t}{t-1},\,\frac{\ddot t}{\dot t} +\frac{\alpha}{1-\alpha} \left(\frac{\dot t}{t+1} + \frac{\dot t}{t-1}\right)\right),
\end{equation}
to obtain the components of $x=x(\tau)$.  This scheme is readily
carried~out, either numerically or (for certain~$\alpha$) symbolically.
Thus one can, \emph{inter alia}, confirm the results of
\myciteasnoun{Ble2013} on the global dynamics of $\alpha=\beta$
May--Leonard systems.

\begin{example}
Suppose $\alpha=m/(m+1)$, for $m$ a positive integer.  Then the exponent
${\alpha/(\alpha-1)}$ in~(\ref{eq:MLode}) equals~$-m$, and ${\rm
  d}\tau/{\rm d}t=\dot t^{-1}$ is a degree\nobreakdash-$3m$ polynomial
in~$t$.  Hence $\tau$~is a degree-$(3m+\nobreak1)$ polynomial, and its
inverse $t=t(\tau)$ is algebraic.  By substituting this function
into~(\ref{eq:MLx}) one deduces that on the complex
$\tau$\nobreakdash-plane, each of $x_1,x_2,x_3$ is also an algebraic
(finite-branched) function.

Hence when $\alpha=\beta=m/(m+1)$, the May--Leonard system has a weak form
of the Painlev\'e property.  This result appears to be new.  The simplest
case is when $\alpha=1/2$ (corresponding to $m=1$), when $\tau$~is quartic
in~$t$.  By applying the quartic formula one can express $t=t(\tau)$ and
hence each $x_i=x_i(\tau)$ in~terms of radicals.
\end{example}

\begin{example}
\label{ex:lambda0}
Let $\alpha=\beta=-1$, so that $(a_1,a_2;b_1,b_2;n)=(1/2,1/2;1,1;1)$.  When
$\alpha=-1$ the solution of the ODE (\ref{eq:MLode}) can be expressed using
the Weierstrassian elliptic function $\wp(\tau)=\wp(\tau;g_2,g_3)$, with
$\dot\wp^2 =\allowbreak 4\wp^3-\nobreak g_2\wp-\nobreak g_3$.  One notes
that the gSE~(\ref{eq:gSE}), also satisfied by $t=t(\tau)$, is an ODE
listed by \myciteasnoun{CartonLeBrun69c}, as her $n=1$ case XIII.4.
Regardless of which ODE one integrates, one finds (with $C$ a constant of
integration) the general solution
\begin{subequations}
\begin{gather}
  t(\tau) = \wp(\tau;\,g_2,g_3) + C, \\
g_2 = 4(3C^2+1), \qquad g_3 = 8C(C^2-1),
\end{gather}
\end{subequations}
where $\tau$~can be replaced by $A\tau+\nobreak B$.  By substituting
into~(\ref{eq:MLx}) and using the identity $\ddot\wp=\allowbreak
6\wp^2-\nobreak g_2/2$ one obtains the general solution $x=x(\tau)$:
\begin{equation}
\label{eq:MLsoln}
  (x_1,x_2,x_3)(\tau) = \left(
\frac{\dot\wp}{\wp+C+1},\,
\frac{\dot\wp}{\wp+C-1},\,
\frac{2\left[(\wp+C)^2-1\right]}{\dot\wp}
\right),
\end{equation}
where $\tau$~can be replaced by $A\tau+\nobreak B$, with $x$~scaled by~$A$.
This being meromorphic on the $\tau$\nobreakdash-plane, the
$\alpha=\beta=-1$ May--Leonard system has the Painlev\'e property.

That this system can be integrated using elliptic functions was noted by
\myciteasnoun{Brenig88}, but the explicit solution~(\ref{eq:MLsoln}) is
new.  It is easily checked that the functions $x_1(x_2-\nobreak 2x_3)$,
$x_2(x_1-\nobreak 2x_3)$ and $x_3(x_1-\nobreak x_2)$ are first integrals:
they are time-independent, depending only on $C$ and~$A$.  They are
essentially the first integrals introduced as $I_1,I_2,I_3$
in~(\ref{eq:strelcyn}) above (where $x$~is to be read as~$y$).  Starting
from~(\ref{eq:MLsoln}), one can confirm that the first integrals found by
\myciteasnoun{Llibre2011} and \myciteasnoun{Tudoran2012} are also
time-independent.

For the general solution $x=x(\tau)$ of a version of the $\alpha=\beta=-1$
May--Leonard system with a single nonzero growth rate, see
example~\ref{ex:portmanteau} below.
\end{example}

\section{Unequal growth rates and Painlev\'e transcendents}
\label{sec:4}

Section~\ref{sec:3} treated $d=3$ Lotka--Volterra systems classified as
`case~$\textrm{II}_i$' according to table~\ref{tab:3tab}, with zero growth
rates.  (The extension to nonzero but equal growth rates is easy, as was
noted.)  Several $d=3$ systems with unequal growth rates, which have the
Painlev\'e property, will now be completely integrated.  They are
deformations of systems treated in~\S\,\ref{sec:3} but do not fit into the
framework of table~\ref{tab:3tab}.  This is because their first integrals,
based on Darboux polynomials, are not conserved: they depend exponentially
on~$\tau$.  Their solutions $x=x(\tau)$ turn~out to involve Painlev\'e
transcendents, which for Lotka--Volterra systems is a novelty.

Consider the system
\begin{equation}
\label{eq:system4}
\left\{
\begin{aligned}
  \dot x_1 &= x_1\left[-a_1x_1 + (1-a_2)x_2 + x_3\right],\\
  \dot x_2 &= x_2\left[+(1-a_1)x_1  -a_2 x_2 + x_3\right],\\
  \dot x_3 &= x_3\left[\lambda + (b_1-a_1)x_1 + (b_2-a_2)x_2 -
    x_3\right],
\end{aligned}
\right.
\end{equation}
which is the $n=1$ specialization of the case-$\textrm{II}_3$ system
(\ref{eq:system3}), modified to include a growth rate~$\lambda$ for
species~3.  Up~to scaling, this is the generic system in which species 1,2
have equal effects on species~3, species~3 has an equal but negative effect
on itself, and species 1,2 have zero growth rates.  As in \S\S\,\ref{sec:2}
and~\ref{sec:3}, define an auxiliary variable~$t$ by $t=(x_1+\nobreak
x_2)/\allowbreak(x_2-\nobreak x_1)$.  By calculus one deduces that
\begin{equation}
  \left(x_1,x_2,x_3\right)(\tau) =  \left(\tfrac{\dot t}{t+1},\,\tfrac{\dot t}{t-1},\,\tfrac{\ddot t}{\dot t} - (1-a_1)\tfrac{\dot t}{t+1} - (1-a_2)\tfrac{\dot t}{t-1}\right)(\tau),
\label{eq:big4}
\end{equation}
which previously appeared as Eq.~(\ref{eq:big2}).

The new time $t=t(\tau)$ satisfies an extended version of the generalized
Schwarzian equation~(\ref{eq:gSE}).  If the left side of~(\ref{eq:gSE}) is
abbreviated as $\text{gS}(a_1,\nobreak a_2;\allowbreak b_1,\nobreak
b_2;n)$, by tedious elimination one deduces that
\begin{equation}
\label{eq:gSEextended}
  \textrm{gS}(a_1,a_2;b_1,b_2;1) + \lambda
\left[
\left(
\frac{1-a_1}{t+1} + \frac{1-a_2}{t-1}
\right)\frac1{\dot t} - \frac1{\dot t^3}
\right]
 = 0.
\end{equation}
It is straightforward to rewrite this nonlinear third-order ODE in the form
$\dot J-\nobreak \lambda J=0$, where
\begin{equation}
  J\defeq
\frac{ \ddot t - \left(\frac{1-a_1}{t+1} + \frac{1-a_2}{t-1} \right) \dot
  t^2 }{(t+1)^{1-2a_1+b_1}(t-1)^{1-2a_2+b_2}}
\end{equation}
can be viewed as a time-dependent first integral.  The existence of such a
representation is unsurprising: by exploiting~(\ref{eq:big4}), one can see
that $J$~is the familiar DP-based first integral~(\ref{eq:firstintegral3})
of~\S\,\ref{sec:3}, written in~terms of $t,\dot t,\ddot t$ (and with
absolute value signs omitted, so that in~general, branch choices must be
made).

To deal with the exponential dependence on~$\tau$, introduce a new
independent variable $z\defeq {\rm e}^{\lambda\tau/c}$ for some $c\neq0$, and
denote ${\rm d}/{\rm d}z$ by a prime.  The statement $J=J_0{\rm e}^{\lambda\tau}$
can be written as a nonlinear second-order ODE for $t=t(z)$, i.e.
\begin{equation}
\label{eq:proto}
  t'' = \left(
\frac{1-a_1}{t+1} + \frac{1-a_2}{t-1}
\right)(t')^2 - \frac{t'}z + K\,\frac{(t+1)^{1-2a_1+b_1}(t-1)^{1-2a_2+b_2}}{z^{2-c}},
\end{equation}
where $K=c^2J_0/\lambda^2$ is a constant of integration.
Equation~(\ref{eq:proto}) is a `proto-Painlev\'e' ODE, in that for certain
$a_1,\nobreak a_2;\allowbreak b_1,\nobreak b_2;c$, it defines a Painlev\'e
transcendent \mycite{Ince27}.  This is obscured by its singular points
being $t=-1,\infty,+1$, while the defining ODE's for the transcendents
(traditional; due to Painlev\'e and Gambier) have singular points
$w=0,1,\infty$.  An equivalent ODE is
\begin{equation}
\label{eq:protow}
  w'' = \left(
\frac{1-a_1}{w} + \frac{1-a_3}{w-1}
\right)(w')^2 - \frac{w'}z - K\,\frac{w^{1-2a_1+b_1}[(w-1)/2]^{1-2a_3+b_3}}{z^{2-c}},
\end{equation}
where $w=\allowbreak (t+\nobreak 1)/\allowbreak (t-\nobreak 1)=x_2/x_1$ is
a new dependent variable, with $t=\allowbreak (w+\nobreak 1)/\allowbreak
(w-\nobreak 1)$; and by definition $a_3\defeq\allowbreak 1-\nobreak
a_1-\nobreak a_2$ and $b_3\defeq\allowbreak 1-\nobreak b_1-\nobreak b_2$.
\begin{example}
\label{ex:portmanteau}
  For certain $(a_1,\nobreak a_2;\allowbreak b_1,\nobreak b_2;c)$, the ODE
  (\ref{eq:protow}) for $w=w(z)$ is or is reducible to a
  Painlev\'e\nobreakdash-III or Painlev\'e\nobreakdash-V equation.  The
  ${\textrm{P}}_{\textrm {III}}$ and ${\textrm{P}}_{\textrm V}$ equations
  have parameters $\alpha,\beta,\gamma,\delta$; and for ${\textrm
    N}={\textrm{III}}$ or~${\textrm V}$,
  $w_{\textrm{N}}(z)=w_{\textrm{N}}(\alpha,\beta,\gamma,\delta;z)$ will
  denote any solution of the respective equation.  The possibilities
  include
  \begin{displaymath}
    \begin{alignedat}{4}
      &(a_1,a_2;\,b_1,b_2;\,c)&&=(0,0;\,\tfrac2r,-\tfrac2r;\,2)&&:&\qquad& w=w_{\textrm{III}}(0,0,K',0;\,z)^r,\\[-1.25pt]
     &\qquad&&=(0,0;\,-\tfrac23,-\tfrac13;\,\tfrac43)&&:&\qquad& w=z^{-1}w_{\textrm{III}}(K',0,0,-K';\,z)^3,\\[-1.25pt]
     &\quad&&=(0,0;\,-\tfrac12,-\tfrac12;\,1)&&:&\qquad& w=w_{\textrm{III}}(K',K',0,0;\,z)^2,\\[-1.25pt]
     &\qquad&&=(0,\tfrac12;\,-\tfrac12,1;\,1)&&:&\qquad& w=R[w_{\textrm{III}}(K',K',0,0;\,z)^2],\\[-1.25pt]
     &\qquad&&=(\tfrac12,\tfrac12;\,1,1;\,1)&&:&\qquad& w=w_{\textrm{V}}({0,0,K',0};z),
    \end{alignedat}
  \end{displaymath}
  where $r\neq0$ is arbitrary, $R[w]\defeq -4w/(w-1)^2$; and $K'\propto K$
  is free in each case.  These are Carton-LeBrun's $n=1$ cases e.IV, XI.1,
  XIII.1, XIII.3, XIII.4.  By substituting $t=\allowbreak (w+\nobreak
  1)/\allowbreak (w-\nobreak 1)$ into~(\ref{eq:big4}), one obtains
  $(x_1,x_2,x_3)$ in~terms of the transcendent $w_{\textrm{N}}(z={\rm e}^{\lambda
    \tau/c})$ and its derivatives.  

  It should be noted that the last of these complete integrations, in~terms
  of~$w_{\textrm V}$, is of a deformed ($\lambda\neq0$) version of the
  $\alpha=\beta=-1$ May--Leonard system, which was solved (with
  $\lambda=0$) in example~\ref{ex:lambda0}.
\end{example}

The preceding was stimulated by later \citeyearpar{CartonLeBrun70} results
of \citeauthor{CartonLeBrun70}, who studied and integrated many gSE's of an
extended type that includes Eq.~(\ref{eq:gSEextended}).

\section{Summary and final remarks}
\label{sec:6}

We have shown [in Eq.~(\ref{eq:parametric3})] how to construct the general
solution $x=x(\tau)$ of any $d=3$ Lotka--Volterra system in which species
$j,k$ have equal effects on species~$i$, and the three species have equal
growth rates.  Such systems had not previously been integrated, despite
extensive work.  The constructed solution is parametric, with $\tau,x$
expressed with the aid of the incomplete beta function as functions of the
new time variable $t\defeq\allowbreak (x_j+\nobreak x_k)/(x_k-\nobreak
x_j)$.  If the system has the Painlev\'e property, the new time is
typically not needed: as a function of~$\tau$, the system state $x$~can be
expressed in~terms of elementary or elliptic functions.

From the complete integration of any Lotka--Volterra system of this type,
one can derive first integrals.  One is the DP-based first integral
of~(\ref{eq:1stintegral3}), the constancy of which defines type~${\rm
  II}_i$.  An additional first integral involving a quadrature was found in
the ABC case by \myciteasnoun{Goriely92}, and more generally by
\myciteasnoun[theorem~5]{Gao2000}.  But they did not construct general
solutions. It seems difficult to go from a pair of first integrals to a
general solution, rather than the reverse.

The fully symmetric ($\alpha=\beta$) May--Leonard model was integrated, and
the $\alpha=\beta=-1$ case was treated in examples \ref{ex:lambda0}
and~\ref{ex:portmanteau}.  If the growth rates are equal, $x$~is expressed
in~terms of the Weierstrassian elliptic function $\wp(\tau)$; and if a
single rate ($\lambda$) is nonzero, the Painlev\'e transcendent $w_{\rm
  V}({\rm e}^{\lambda\tau})$.  This case, with equal growth rates, was
studied geometrically by \myciteasnoun{Tudoran2012}, but the elliptic
general solution is new.  The appearance of a Painlev\'e function is also
novel.

The results of \myciteasnoun{CartonLeBrun69c,CartonLeBrun70} on certain
nonlinear third-order ODE's with the Painlev\'e property proved invaluable.
They are not well known, though they have been re-worked by
\myciteasnoun[Appendix~C]{Cosgrove97}.  In many of our examples they
facilitated the integration of the ODE satisfied by the new time
$t=t(\tau)$.  With further effort, they could probably be made to yield a
classification of all type\nobreakdash-${\rm II}_i$ Lotka--Volterra systems
with the Painlev\'e property.

Our techniques may be useful in constructing solutions of other
small\nobreakdash-$d$ quadratic dynamical systems, of the sort reviewed in
the introduction.  A related inverse problem is also of interest.  Many
functions of time~$\tau$ have Lotka--Volterra representations, i.e., can be
generated from solutions $x=x(\tau)$ of systems of Lotka--Volterra type.
(This is stressed by \myciteasnoun{Peschel86}, who even mention hardware
implementations.)  Painlev\'e functions are included, as \S\,\ref{sec:4}
made clear.  Along this line, the quadratic representability of solutions
of further higher-order ODE's of Painlev\'e type, such as Chazy equations,
will be explored elsewhere.

\appendix

\section*{Appendix: Incomplete beta functions and their inverses}

In a normalization used here, the incomplete beta function
${\mathrm B}_{{a}_1,{a}_2;t_0}(t)$, for (real) indices ${a}_1,{a}_2$,
(real) endpoints $-1,1$ and a (real) basepoint $t_0\neq -1,1$, is given by
\begin{equation}
  {\mathrm B}_{{a}_1,{a}_2;t_0}(t)\defeq \int_{t_0}^t |t'+1|^{{a}_1-1}\,
  |t'-1|^{{a}_2-1}\,dt'.
\label{eq:a1}
\end{equation}
The basepoint $t_0$ lies in one of the intervals $(-\infty,-1)$, $(-1,1)$,
$(1,\infty)$, and by convention $\tau={\mathrm B}_{{a}_1,{a}_2;t_0}(t)$ is
defined only for~$t$ in that interval.  It is an increasing function
of~$t$.  If ${a}_1,{a}_2$ are positive integers it is a polynomial
function; and if $a_1,a_2$ are integers with $a_1a_2<0$ and $a_1+\nobreak
a_2\le0$, it is a rational function.  In~general it is a `special' (higher
transcendental) function.  If $a_1,a_2>0$ it can be expressed in~terms of
the traditionally normalized incomplete beta function, which is
\begin{equation}
  \label{eq:Bdef}
  {\mathrm B}_{\hat t}(a_1,a_2) = \int_0^{\hat t} t^{a_1-1}(1-t)^{a_2-1}\,dt
\end{equation}
for $\hat t\in(0,1)$.  Specifically,
\begin{equation}
  2^{1-a_1-a_2}\,{\mathrm B}_{a_1,a_2;t_0}(t) =
  \begin{cases}
    {\mathrm B}_{(t+1)/(t-1)}\,(a_1,1-a_1-a_2) + C, & t_0\in(-\infty,-1), \\
    {\mathrm B}_{(t+1)/2}\,(a_1,a_2) + C, & t_0\in(-1,1), \\
    {\mathrm B}_{(t-1)/(t+1)}\,(a_2,1-a_1-a_2) + C, & t_0\in(1,\infty),
  \end{cases}
\end{equation}
with $C=C_{a_1,a_2;t_0}$ chosen so that ${\mathrm B}_{a_1,a_2;t_0}(t_0)=0$.
The traditional function ${\mathrm B}_{\hat t}(a_1,a_2)$ is supported by
many software packages, and many identities, expansions and approximations
for~it are known \mycite{Dutka81}.  In particular,
\begin{equation}
{\mathrm B}_{\hat t}(a_1,a_2) = 
a_{1}^{-1} \,{\hat t}^{a_1}(1-\hat t)^{a_2}\,
{}_2F_1(a_1+a_2,1;\,a_1+1;\, \hat t),
\end{equation}
where ${}_2F_1$ is the Gauss hypergeometric function.  Thus for many
choices of $a_1,a_2$, closed-form expressions exist
\mycite[\S\,7.3]{Prudnikov86c}.  A few have been derived in the context of
hyperlogistic population growth \mycite{Blumberg68}.

The \emph{inverse} incomplete beta function $t={\mathrm
  B}^{-1}_{{a}_1,{a}_2;t_0}(\tau)$ is also an increasing function of its
argument, and is algebraic if ${a}_1,{a}_2$ are positive integers, or
integers with $a_1a_2<0$ and $a_1+\nobreak a_2\le0$.  This inverse function
is defined on some interval containing $\tau=0$, and satisfies (with $\dot
t={\rm d}t/{\rm d}\tau$) the hyperlogistic growth equation
\begin{equation}
\label{eq:appeq}
  \dot t = |t+1|^{1-{a}_1}|t-1|^{1-{a}_2},
\end{equation}
with the initial condition $t(\tau=0)=B^{-1}_{{a}_1,{a}_2;t_0}(0)=t_0$.  It
maps the $\tau$\nobreakdash-interval onto whichever of $(-\infty,-1)$,
$(-1,1)$, $(1,\infty)$ contains~$t_0$.  It is useful to write
\begin{equation}
\label{eq:appeq2}
  {\mathrm B}^{-1}_{{a}_1,{a}_2;t_0} (\tau)  = {\mathrm B}^{-1}_{{a}_1,{a}_2}(\tau-\tau_0),
\end{equation}
where $t={\mathrm B}^{-1}_{{a}_1,{a}_2}(\tau)$ is any convenient,
standardized solution of~(\ref{eq:appeq}) that maps \emph{some}
$\tau$\nobreakdash-interval $(\tau_\textrm{min},\tau_\textrm{max})$ onto
the $t$\nobreakdash-interval containing~$t_0$.  The time-origin~$\tau_0$ is
determined by the condition that ${\mathrm
  B}_{{a}_1,{a}_2}^{-1}(-\tau_0)=t_0$.

The function $\tau={\mathrm B}_{{a}_1,{a}_2;t_0}(t)$ can be continued
analytically from its real interval of definition to the upper half of the
complex $t$\nobreakdash-plane.  (The continuation is given by~(\ref{eq:a1})
without absolute value signs, multiplied by an overall phase factor.)  The
continuation is a Schwarzian triangle function \mycite{Nehari52} that
performs a Schwarz--Christoffel transformation: provided
${a}_1,{a}_2,\allowbreak 1-\nobreak{a}_1-\nobreak {a}_2$ are non-negative,
it conformally maps the upper half of the complex $t$\nobreakdash-plane to
the interior of some triangle $\Delta ABC$ in the $\tau$\nobreakdash-plane
with vertex angles $\pi({a}_1,{a}_2,\allowbreak
1-\nobreak{a}_1-\nobreak{a}_2)$, the points $-1,1,\infty$ on the real
$t$\nobreakdash-axis being taken to the vertices $A,B,C$.  Many conformal
mapping functions of this type, which are essentially incomplete beta
functions, can be found in the catalogue of \myciteasnoun{vonKoppenfels59}.
In each case the inverse function takes the interior of $\Delta ABC$ to the
upper-half $t$\nobreakdash-plane.

This inverse $t={\mathrm B}^{-1}_{{a}_1,{a}_2;t_0}(\tau)$ can sometimes be
given in closed form; e.g., when the unordered set
$\{1/{a}_1,1/{a}_2,\allowbreak 1/(1-\nobreak{a}_1-\nobreak{a}_2)\}$ is any
of $\{2,4,4\}$, $\{2,3,6\}$ or $\{3,3,3\}$.  In these cases the inverse can
be continued to the entire complex $\tau$\nobreakdash-plane from its real
interval of definition, and furthermore from $\Delta ABC$, by applying the
Schwarz reflection principle: reflecting repeatedly through the sides of
the triangle.  The resulting one-valued function of~$\tau$ is elliptic
(i.e.\ doubly periodic) and can be given explicitly.  The cases
$\{1,m,-m\}$ (for $m$ any positive integer), $\{1,\infty,\infty\}$ and
$\{2,2,\infty\}$ are similar, but yield inverse functions $t={\mathrm
  B}_{{a}_1,{a}_2;t_0}^{-1}(\tau)$ that are elementary rather than
elliptic.

\begin{table}
  \caption{Five cases when $t=t(\tau)={\mathrm B}_{{a_1},{a_2}}^{-1}(\tau)$ is
    expressible in closed form and can be extended to a one-valued function
    on the complex $\tau$\nobreakdash-plane.}
  \begin{center}
    {
      \small
      \begin{tabular}{l||ll|ll}
        \hline
        ${a}$ & \multicolumn{4}{c}{$t=t(\tau)={\mathrm B}_{{a},{a}}^{-1}(\tau)$, \quad $\tau\in(\tau_\textrm{min},\tau_\textrm{max})$} \\
        \hline
        & \multicolumn{2}{c}{$t_0\in(-1,1)$ subcase} & \multicolumn{2}{c}{$\pm t_0 \in(1,\infty)$ subcase} \\
        \hline
        \hline
        $0$ &  $\tanh\tau$, & $\tau\in(-\infty,\infty)$ & $-\coth\tau$, & $\pm\tau\in(-\infty,0)$ \\
        $\frac14$ & $\sqrt2\, [{\rm sn}\,{\rm dn}](\tau/\sqrt2)$, &
        $\tau\in(-\bar K_{1/4},\bar K_{1/4})$ & $\pm\frac12[{\rm cn}^2+{\rm cn}^{-2}](\tau/2)$, & $\pm\tau\in(0, K_{1/4})$\\
        $\frac13$ & $-81\,\dot\wp/[2\,(9\,\wp+1)^2]$, & $\tau\in(-\bar K_{1/3},\bar K_{1/3})$ & $\pm1 - 8/(27\,\dot\wp\pm2)$, & $\pm\tau\in(0, K_{1/3})$ \\
        $\frac12$ & $\sin\tau$, & $\tau\in(-\pi/2,\pi/2)$ & $\pm\cosh\tau$,  & $\pm\tau\in(0,\infty)$  \\
        $1$ &  $\tau$, & $\tau\in(-1,1)$ & $\pm1+\tau$, & $\pm\tau\in(0,\infty)$ \\
        \hline
      \end{tabular}
    }
  \end{center}
  \label{tab:3}
\end{table}

Thus there are many `conformally distinguished' choices for the unordered
set $\{1/{a}_1,1/{a}_2\}$, each of which yields an explicit formula for the
inverse function $t={\mathrm B}^{-1}_{{a}_1,{a}_2;t_0}(\tau)$.  The elementary cases
are $\{1,m\}$, $\{1,-m\}$, $\{m,-m\}$, $\{1,\infty\}$, $\{2,2\}$,
$\{2,\infty\}$, $\{\infty,\infty\}$, and the elliptic ones are $\{2,3\}$,
$\{2,4\}$, $\{2,6\}$, $\{3,3\}$, $\{3,6\}$, $\{4,4\}$.  In each of these
cases any standardized inverse function $t={\mathrm B}^{-1}_{{a}_1,{a}_2}(\tau)$ of
the type used in~(\ref{eq:appeq2}), which has a real interval of definition
$(\tau_\textrm{min},\tau_\textrm{max})$, can optionally be continued to a
one-valued function on the $\tau$\nobreakdash-plane.

Table~\ref{tab:3} gives such a standardized function $t=t(\tau)={\mathrm
  B}^{-1}_{{a},{a}}(\tau)$ for ${a}=0,\frac14,\frac13,\frac12,1$.  These
are the conformally distinguished cases with ${a}_1={a}_2\eqdef{a}$, and
are of particular interest (the ones with ${a}_1\neq{a}_2$ are left to the
reader).  In each case there are two subcases: $t_0\in(-1,1)$ and $\pm
t_0\in(1,\infty)$.  The function $t=t(\tau)$ maps
$\tau\in\allowbreak(\tau_\textrm{min},\tau_\textrm{max})$ onto
$t\in(-1,1)$, resp.\ $\pm t\in(1,\infty)$, and by~(\ref{eq:appeq}) it
satisfies
\begin{equation}
\dot t=  (1-  t^2)^{1-{a}},\qquad \textrm{resp.}\qquad
\dot t = (t^2-   1)^{1-{a}}.
\label{eq:appodes}
\end{equation}
In the table each $t=t(\tau)$ has been chosen to satisfy $t(\tau=0)=0$,
resp.\ $t(\tau=0)=\pm1$ (except when ${a}=0$).  Each is increasing on its
interval of definition $(\tau_\textrm{min},\tau_\textrm{max})$.

The elementary functions appearing in the table (for ${a}=0,\frac12,1$)
follow by inspection: by integrating the ODE's~(\ref{eq:appodes}).  The
elliptic solutions of these ODE's (for ${a}=\frac14,\frac13$) are less
obvious, but can readily be derived with the aid of elliptic function
identities from functions that are known to map certain triangles
conformally to the upper half-plane \mycite{Kober57,Sansone69}.  For
${a}=\frac14$ the Jacobian elliptic functions ${\rm sn},{\rm cn},{\rm dn}$
that appear are `lemniscatic': their common modular parameter $m=k^2$
equals~$\frac12$.  For ${a}=\frac13$ the Weierstrassian elliptic function
$\wp(\tau)=\wp(\tau;g_2,g_3)$ that appears is `equianharmonic': its
parameter~$g_2$ equals zero.  Also, its parameter~$g_3$ equals $-4/3^6
=\allowbreak -4/729$.  It satisfies the usual Weierstrassian ODE $\dot\wp^2
=\allowbreak 4\wp^3-\nobreak g_2\wp-\nobreak g_3$, i.e., $\dot\wp^2
=\allowbreak 4(\wp^3+\nobreak 3^{-6})
$.

For each function $t=t(\tau)={\mathrm B}^{-1}_{{a},{a}}(\tau)$ appearing in the
table, the interval of definition
$\tau\in(\tau_\textrm{min},\tau_\textrm{max})$ is easily computed
from~(\ref{eq:a1}).  It can be expressed in~terms of the traditionally
normalized \emph{complete} beta function
$\mathrm{B}({a}_1,{a}_2)\defeq\allowbreak
\Gamma({a}_1)\Gamma({a}_2)/\,\allowbreak \Gamma({a}_1+\nobreak {a}_2)$.  In
the subcases $t_0\in(-1,1)$ and $\pm t_0\in(1,\infty)$, it is
\begin{equation}
  \tau\in (-\bar K_{a}, \bar K_{a}), \qquad \textrm{resp.} \qquad
  \pm\tau\in (0,  K_{a}),
\end{equation}
where
\begin{displaymath}
  \bar K_{a} \defeq \tfrac12\, \mathrm{B}({a},\tfrac12) = \frac12\,
  \frac{\Gamma({a})\,\Gamma(\tfrac12)}{\Gamma({a}+\frac12)}, \quad
  \textrm{resp.} \quad K_{a} \defeq \tfrac12\,
  \mathrm{B}({a},\tfrac12-{a}) = \frac12\,
  \frac{\Gamma({a})\,\Gamma(\frac12-{a})}{\Gamma(\frac12)}.
\end{displaymath}
It is understood that $\bar K_{a}$ is infinite if ${a}\le0$.  For instance,
$\bar K_0, \bar K_{1/2}, \bar K_1$ are $\infty,\pi/2,1$, as shown.  Also,
in the $\pm t_0\in(1,\infty)$ case the interval $\pm\tau\in(0,K_a)$ must be
replaced by $\pm\tau\in(-\infty,0)$ if $a\le0$, and interpreted as
$\pm\tau\in(0,\infty)$ if $a\ge\frac12$.

The values of~$a$ in the table are not the only ones for which the inverse
function $t=t(\tau)={\mathrm B}^{-1}_{a,a}(\tau)$ can be expressed in
closed form.  It is an algebraic function if $a$~is any positive integer,
and if $a=2$ it can even be expressed in~terms of radicals.  But for any
integer $a>1$, its continuation to the $\tau$\nobreakdash-plane is
multiple-valued.

No similar table of closed-form expressions seems to have appeared
previously; though certain values of the traditionally normalized inverse
incomplete beta function (especially for large (half-)integral values
of~$a_1,a_2$) were calculated and tabulated long ago, owing to their
importance in statistics \mycite{Thompson41}.

\small

\setlength\bibsep{-0.3pt}
\makeatletter
   \@setfontsize\small\@ixpt{10.75}
\makeatother



\end{document}